\documentclass[12pt]{article}
\usepackage{amsthm,amsmath}
\usepackage{amsfonts}
\usepackage{setspace}
\usepackage[margin=1.0in]{geometry}
\usepackage{natbib}
\usepackage{tikz}
\usepackage{cleveref}
\usepackage{amsbsy}
\usepackage[shortlabels]{enumitem}
\usepackage{bm}
\usepackage{algorithm} 
\usepackage{algpseudocode} 
\usepackage{float}
\usepackage{comment}
\usepackage{thmtools}
\usepackage{thm-restate}
\usepackage{xcolor}
\usepackage{footnote}
\usepackage{sidecap}

\makesavenoteenv{tabular}
\makesavenoteenv{table}

\DeclareMathOperator*{\argmin}{arg\,min}

\usepackage{soul}
\newcommand{\nb}[3]{{\colorbox{#2}{\bfseries\sffamily\scriptsize\textcolor{white}{#1}}}
 {\textcolor{#2}{\sf\small\textit{#3}}}}

\newcommand{\na}[1]{\nb{Nick}{blue}{#1}}

\newcommand{\todo}[1]{\nb{TODO}{red}{#1}}

\newcommand{\E}[1]{\mathbb{E}\left[ #1 \right]}
\newcommand{\PP}[1]{\mathbb{P}\left( #1 \right)}

\newcommand{\offopt}[0]{\textup{OMN}(\mathcal{I})}
\newcommand{\offoptt}[0]{\textup{OMN}(\mathcal{I},t)}
\newcommand{\offrb}[0]{\textup{LP}^{\textup{OMN}}(\mathcal{I})}
\newcommand{\offrbrel}[0]{\textup{LP}^{\textup{OMN}}_{\textup{REL}}(\mathcal{I})}

\newcommand{\offrdrel}[0]{\textup{DLP}^{\textup{OMN}}_{\textup{REL}}(\mathcal{I})}

\newcommand{\alg}[0]{\textup{ALG}(\mathcal{I})}
\newcommand{\plp}[0]{\textup{LP}^{\textup{ALG}}}
\newcommand{\plpm}[0]{\textup{LP}^{\textup{ALG}}(\mathcal{I, \m{M}})}
\newcommand{\plpt}[0]{\textup{LP}^{\textup{ALG}}(\mathcal{I, \m{T} \times \m{T}})}

\newcommand{\dlplong}[0]{\textup{DLP}^{\textup{ALG}}(\mathcal{I}, \m{T} \times \m{T})}
\newcommand{\dlpm}[0]{\textup{DLP}^{\textup{ALG}}(\mathcal{I}, \m{M})}

\newcommand{\bx}{{\bf x}}

\newcommand{\m}[1]{\mathcal{#1}}
\newcommand{\prefixes}[1]{\text{Prefixes}(#1)}

\theoremstyle{plain}

\newtheorem{definition}{Definition}
\newtheorem{theorem}{Theorem}

\newtheorem{lemma}{Lemma}
\newtheorem{corollary}{Corollary}
\newtheorem{proposition}{Proposition}
\newtheorem{conjecture}{Conjecture}

\newtheorem*{theorem*}{Theorem}
\newtheorem{example}{Example}

\title{Greedy Dynamic Matching}
\author{Nick Arnosti, Felipe Simon}
\date{}
\begin{document}
\maketitle

\begin{abstract}
	We study a foundational model of dynamic matching market with abandonment. This model has been studied by \citet{collina2020dynamic} and \citet{aouad2022dynamic}, and many other papers have considered special cases. 
	We compare the performance of greedy policies -- which identify a set of ``acceptable'' matches up front, and perform these matches as soon as possible -- to that of an omniscient benchmark which knows the full arrival and departure sequence.
	
	We use a novel family of linear programs ($\textup{LP}^\textup{ALG}$) to identify which greedy policy to follow. 
	We show that the value of $\textup{LP}^\textup{ALG}$ is a {\em lower bound} on the value of the greedy policy that it identifies in two settings of interest:
	\begin{itemize}
	\item Everyone has the same departure rate. \vspace{-.01 in}
	\item Bipartite matching where everyone on the same side of the market has the same departure rate.
	\end{itemize}
	Our proofs use a new result (\Cref{prop:gammabound}), which provides a lower bound on the {\em probability} that at least one agent from a set of types is present, in terms of the {\em expected number} of such agents.
	 
	We also show that the value of $\textup{LP}^\textup{ALG}$ is at least 1/2 of the reward rate earned by the omniscient policy (\Cref{prop:factor2}). Therefore, for both settings above, our greedy policy provably earns at least half of the omniscient reward rate. This improves upon the bound of $1/8$ from \citet{collina2020dynamic}. In both settings our competitive ratio of $1/2$ is the best possible: no online policy can provide a better guarantee (\Cref{thm:no-better-half}).
	
	Numerical evidence suggests that our algorithm achieves $1/2$ of the omniscient benchmark even when our departure rate conditions are not satisfied. We also establish that our algorithm achieves $1/2$ for the case with two types (\Cref{thm:two-types}).
		
	To show these results we introduce a new linear program that upper bounds the objective value of the omniscient policy (\Cref{prop:offline-bound}). This improves upon the upper bounds presented by \citet{collina2020dynamic} and \citet{kessel2022stationary}.

\end{abstract}

\newpage
\tableofcontents

\newpage
\section{Introduction}

We study a centralized dynamic matching market with pairwise matching of impatient agents. 
In our model, there is a finite set of agent types $\m{T}$. Agents of type $i \in \m{T}$ arrive according to a Poisson process of rate $\lambda_i$. Agents depart the system when matched to another agent, but can also {\em abandon} the system without matching. The maximum time that an agent of type $i$ will remain in the system before abandoning is drawn from an exponential distribution with rate $\mu_i$. The reward of matching an agent of type $j$ with an agent of type $i$ who arrived earlier is $r_{ij}$. An {\em instance} $\m{I}$ of our problem is fully specified by types $\m{T}$, arrival rates $\{\lambda_i\}_{i \in \m{T}}$, abandonment rates $\{\mu_i\}_{i \in \m{T}}$, and rewards $\{r_{ij}\}_{i,j \in \m{T}}$. 

A central planner must choose a policy $\pi$, which specifies which agents to match. We are interested in the long-run (steady-state) rate of reward, which we denote $V(\pi, \mathcal{I})$ and define precisely in \Cref{sec:offlineoptimal}. Our benchmark is the maximum reward rate that an omniscient policy (with perfect information about future arrivals and departures) can obtain. 

Dynamic matching with abandonment arises across diverse application domains including kidney donation, public housing, and online gaming. As such, our problem has been studied by \citet{collina2020dynamic} and \citet{aouad2022dynamic} (though the latter uses the optimal {\em online} policy as a benchmark). Special cases of this problem, where the graph of possible matches is bipartite and one side must be matched immediately, have been studied by  \citet{kessel2022stationary} and \citet{patel2024combinatorial}.




We study the performance of ``greedy'' policies, which identify a set of acceptable matches, and make these matches as soon as they become available. More precisely, a greedy policy is specified by ``preferences'' $\succ_j$ for each type $j \in \m{T}$. Each $\succ_j$ is a ranking of a subset of $\m{T}$; we refer to this subset as the ``acceptable'' matches for $j$. Given these preferences, a greedy policy operates as follows. When an agent of type $j$ arrives, it is matched with its most preferred type that is present. If no acceptable types are present, then this agent joins the pool of waiting agents. 

{\bf Our contributions.} 
We provide an algorithm which, given an instance $\mathcal{I}$, recommends a greedy policy $\alg$. We compare the performance of our algorithm to an omniscient benchmark which has perfect information about arrivals and departures. We let $\offoptt$ denote the expected reward rate of this benchmark when matches are made on the finite time interval $[0,t]$ and define $\offopt  = \lim_{t\to \infty} \offoptt$. Our main result establishes conditions under which our recommended greedy policy achieves at least half of $\offopt$.

The following definitions capture the conditions required for this result to hold.

\begin{definition} \label{def:homogeneous}
    Instance $\m{I}$ is {\bf bipartite} if the set of types can be partitioned into $\m{T}_1$ and $\m{T}_2$ such that $r_{ij} \le 0$ if $\{i,j\} \subseteq \m{T}_1$ or $\{i,j\} \subseteq \m{T}_2$. 
\end{definition}
\begin{definition}
    Instance $\m{I}$ has {\bf homogenous departures} if either
    \begin{itemize}
        \item $\mu_i = \mu_j$ for all $i,j \in \m{T}$
        \item or $\m{I}$ is bipartite and $\mu_i = \mu_j$ for all $\{i,j\} \subseteq \m{T}_k$ for $k\in \{1,2\}$.
    \end{itemize}
\end{definition}

\begin{theorem} \label{thm:emu:result}
For any instance $\m{I}$ with homogenous departures, 
    \[V(\alg, \mathcal{I})\ge \frac{1}{2} \offopt.\]
\end{theorem}

This result improves upon that of \citet{kessel2022stationary}, who show that a competitive ratio of $1/2$ is obtainable when the compatibility graph is bipartite, there is a single type on one side, and the other side abandons immediately. For comparison, \Cref{thm:emu:result} allows an arbitrary number of types on each side, and allows both sides to have some patience.

\Cref{thm:emu:result} also improves upon the competitive ratio of $1/8$ established by \citet{collina2020dynamic}, which does not require the assumption of homogeneous departures. It is possible that the conclusion of \Cref{thm:emu:result} also holds without this assumption: our simulations have yet to identify an instance where $V(\alg, \mathcal{I}) < \frac{1}{2} \offopt$. In \Cref{sec:app:twotypes}, we establish that $\alg$ also guarantees $1/2$ of the offline benchmark in settings {\em without} homogenous departures, when there are only two types.



Our bound of $1/2$ is tight: no online policy can offer a better guarantee. 

\begin{theorem}
Fix $\epsilon > 0$. There exists an instance $\mathcal{I}$ with homogenous departures such that for any Markovian policy $\pi$, 
\[V(\pi, \mathcal{I}) < \left(\frac{1}{2} + \epsilon\right)\offopt.\]
 \label{thm:no-better-half}
\end{theorem}
An upper bound of $1/2$ can be established using a bipartite instance with three types in which all agents have identical departure rates, or using a non-bipartite instance with two types that have identical departure rates. The proof of \Cref{thm:no-better-half} is in \Cref{appendix:no-better-half}. 


There are two components to the  main result (\Cref{thm:emu:result}) of this paper. The first is an algorithm to identify a greedy policy. This involves two steps.
\begin{enumerate}[1.]
    \item {\bf Novel family of linear programs}. We introduce a novel family of linear programs $\plpm$, each associated with a set of possible matches $\m{M} \subseteq \m{T} \times \m{T}$. We show that certain ``suitable'' solutions of $\plpm$ map naturally to greedy policies.
    \item {\bf Finding suitable solutions}. We provide a procedure in \Cref{algo:suitability} that finds a set $\m{M}$ where where the optimal solution of $\plpm$ is suitable. 
\end{enumerate} 

The second component is to prove that our algorithm earns a reward of at least $\frac{1}{2} \offopt$. The proof uses the following steps.
\begin{enumerate}[1.]
    \item {\bf Lower bound for our greedy policy} (\Cref{prop:plplowerbound}). If the instance $\m{I}$ has homogenous departures, then the value of the optimal solution of $\plpm$ is a lower bound for the value of the associated greedy policy. 
    \item {\bf Upper bound for any policy} (\Cref{prop:offline-bound}). We provide a new linear program that upper bounds $\offopt$. This bound is tighter than those used in previous work.
    \item {\bf Relating lower and upper bounds} (\Cref{prop:factor2}). We show that for the $\m{M}$ selected by \Cref{algo:suitability}, the value of $\plpm$ is at least half of the value of the upper bound from \Cref{prop:offline-bound}.
\end{enumerate}



We can summarize the proof with the following sequence of inequalities: 

\[\frac{1}{2} \offrb \stackrel{\text{Prop. \ref{prop:factor2}}}{\le} \plpm \stackrel{\text{Prop. \ref{prop:plplowerbound}}}{\le} V(\alg, \mathcal{I})\le  \offopt   \stackrel{\text{Prop. \ref{prop:offline-bound}}}{\le} \offrb \]

\Cref{prop:offline-bound} and \Cref{prop:factor2} apply to any instance $\m{I}$; only \Cref{prop:plplowerbound} imposes specific conditions on $\m{I}$. Although all three steps of this proof introduce new ideas, we see \Cref{prop:plplowerbound} as our greatest innovation, as it shows that the linear program $\plpm$ serves as a {\em lower} bound on the value of our greedy policy. All prior work of which we are aware only uses linear programs to provide an {\em upper} bound, and then attempts to guarantee a fraction of this upper bound.

The key component to proving \Cref{prop:plplowerbound} is \Cref{prop:gammabound}. Say that an agent is {\em present} at time $t$ if the agent arrived before $t$ and has not abandoned or been matched before $t$. For $i \in \m{T}$, let $N_i^\pi(t)$ be a random variable representing the number of agents of type $i$ that are present at time $t$. We refer to $N^\pi(t) = \{N_i^\pi(t)\}_{i \in \m{T}}$ as the {\em state} of the system at time $t$, and say that policy $\pi$ is {\em Markovian} if the induced $N^\pi$ is a continuous time Markov chain. \Cref{lemma:stationaryctmc} in \Cref{appendix:steady-state-greedy-policies} establishes that any Markovian policy has a unique steady-state distribution.
\begin{lemma}
    \label{prop:gammabound}
    For any Markovian policy $\pi$, let $N^\pi = \{N_i^\pi\}_{i \in \m{T}}$ be sampled from the steady-state distribution induced by $\pi$. Let $S \subseteq \m{T}$, and define
    \begin{equation} \vspace{-.1 in}\gamma_S = \frac{1-e^{-\sum_{i \in S}\lambda_i/\mu_i}}{\sum_{i \in S}\lambda_i/\mu_i}. \label{eq:gammadef} \end{equation} 
    If $\mu_i = \mu_j$ for all $i, j \in S$, then
    \begin{equation}\vspace{-.1 in} \mathbb{P}(\sum_{i \in S}N_i^\pi > 0)\ge \gamma_S  \mathbb{E}[\sum_{i \in S} N_i^\pi]. \label{eq:key-ineq} \end{equation} 
\end{lemma}
We now interpret this result. Say that an agent of any type $i \in S$ is an $S$-agent. Markov's inequality states that the probability that an $S$-agent is present can be upper bounded by the expected number of present $S$-agents:
\begin{equation} \mathbb{P}(\sum_{i \in S}N_i^\pi > 0) \le  \mathbb{E}[\sum_{i \in S} N_i^\pi]. \end{equation}
\Cref{prop:gammabound} can be seen as a complement to Markov's inequality. It shows that multiplying $\mathbb{E}[\sum_{i \in S} N_i^\pi]$ by the constant $\gamma_S$ provides a {\em lower} bound on the probability that an $S$-agent is present. In other words, it is impossible for the probability of an $S$ agent being present to be too small, if the expected number of $S$-agents is large.

For the policy that makes no matches, \eqref{eq:key-ineq} holds with equality for all $S$ (without needing any assumptions on the $\mu_i$). This follows because when no matches are made, the $N_i^\pi$ are independent Poisson random variables with mean $\lambda_i/\mu_i$. Thus, one interpretation of \Cref{prop:gammabound} is that the ratio of the {\em expected} number of $S$-agents in the system to the probability that there is at least {\em one} $S$-agent is maximized when no matches are made.


The remainder of this manuscript is structured as follows. The next section presents a review of related literature. \Cref{sec:model} formalizes the mode and defines omniscient and greedy policies. \Cref{sec:procedure} describes  our procedure to identify the greedy policy $\alg$.  \Cref{sec:results} presents our main theoretical results. \Cref{sec:numerical} uses numerical simulations to test the performance of $\alg$ in settings without homogeneous departures. \Cref{sec:discussion} concludes by discussing future work and associated conjectures.

\subsection{Literature Review} 

A number of recent papers have either studied the same model as we do, or special cases of this model. \Cref{table:summary} summarizes the known results for different settings and benchmarks.

{\bf Dynamic Matching with Abandonment.} \citet{collina2020dynamic} and \citet{aouad2022dynamic} study the same model as we do.  \citet{collina2020dynamic} finds a policy that guarantees $1/8$ of the omniscient policy, while  \citet{aouad2022dynamic}  offer a guarantee of $\frac{1}{4}(1-1/e) \approx .158$ against the optimal {\em online} policy. \citet{li2023fully} study a discrete version of this model with general (non-exponential) sojourn times and under mild conditions on this sojourn time, find an online policy that achieves a competitive ratio of $0.155$.

{\bf Bipartite Matching and Immediate Departures.} Improved guarantees are possible with stronger assumptions on the instance. For example, \citet{aouad2022dynamic} get a guarantee of $\frac{1}{2}(1-1/e)$ (relative to the optimal {\em online} algorithm) if the set of compatible matches forms a bipartite graph, and $(1-1/e)$ if in addition the vertices on one side of the graph depart immediately.
The setting with bipartite compatibility and instantaneous departure for agents on one side is referred to as the {\it Stationary Prophet Inequality} problem by \citet{kessel2022stationary}. They show that if there is only one ``patient'' type, then a greedy policy guarantees  $1/2$  of omniscient.  


\begin{table}[!t]
\begin{tabular}{l | l l }
    
{\bf Setting} & {\bf Omniscient Benchmark} & {\bf Optimal Online Benchmark} \\ \hline
General & $\frac{1}{8}$ \citep{collina2020dynamic}\footnote{The original proof has a mistake but we still believe the result to hold true.}   & $\frac{1}{4} (1-1/e  \textcolor{red}{+\eta})$\citep{aouad2022dynamic}  \vspace{.1 in} \\
Bipartite & $\frac{1}{4} (1-1/e)$ (This Work, \Cref{sec:app:aouadguarantee}) & $\frac{1}{2} (1-1/e  \textcolor{red}{+\eta})$  \citep{aouad2022dynamic} \vspace{.1 in} \\
Prophet & $1 - 1/\sqrt{e}  \textcolor{red}{+\eta}$ \citep{patel2024combinatorial}& $1 - 1/e  \textcolor{red}{+\eta}$ \citep{aouad2022dynamic}\\
Secretary &  \vspace{-.1 in}
\end{tabular}
\caption{A summary of the best known bounds for different settings and benchmarks. All of these bounds are credited to the papers that first introduced them; those market with red $\textcolor{red}{+\eta}$ have recently been infinitesimally improved by \citet{amanihamedani2024improved}. None of these bounds are conjectured to be tight. We conjecture that for the omniscient benchmark, a tight bound of $1/2$ can always be obtained using a well-chosen greedy policy. We prove this conjecture for instances with homogenous departure rates.}
\label{table:summary}
\end{table}



{\bf Greedy Matching: Beyond Competitive Ratios.}
\citet{kerimov2023optimality,kerimov2024dynamic} study a similar model, but without abandonment. They also compare the performance of greedy policies to the omniscient benchmark, but seek a constant {\em additive} regret (rather than a {\em multiplicative} competitive ratio). This stronger guarantee is not achievable by any online policy in the model with abandonment. 

\citet{kohlenberg_2025} studies the performance of greedy policies in a bipartite setting where ``supply'' is matched to ``demand'', and the reward of a match depends only on the type of demand that is served. This work looks for policies that approximately minimize the ``cost of impatience''. Roughly speaking, this means that the loss from unmatched demand under their policy is within a constant factor of what it would be under any policy. Approximately minimizing the cost of impatience and approximately maximizing total rewards are logically distinct goals, and neither implies the other.

{\bf Known Departure Times.} This paper assumes that departure time is not observable. Other work has shown that better guarantees are possible when this assumption is relaxed 
\citet{akbarpour2020thickness} and \citet{huang2018match} study {\em unweighted} matching with known departure times. \citet{akbarpour2020thickness} consider stochastic arrivals and compatibility, whereas \citet{huang2018match} assume an adversarial sequence of arrivals. \citet{akbarpour2020thickness} focus on approximately minimizing the number of unmatched agents (similar to the cost of impatience metric used by \citet{kohlenberg_2025}) and conclude that knowing when agents are about to abandon allows the planner to substantially decrease the number of unmatched agents. 

\citet{huang2018match, huang2019tight, huang_2020} also study the performance of online policies against an omniscient benchmark but using a different model. They assume that all matches have the same value, but allow arrivals and departures to be adversarial. Crucially, however, the planner gets an opportunity to match vertices just before they abandon. In bipartite problems, an online policy can guarantee a competitive ratio of $0.569$. They also find a $0.521$ guarantee for the general problem. The fact that they can achieve guarantees greater that $0.5$, which are impossible in our setting, points to the value of having precise departure information. 

{\bf Our Contributions.} One of the biggest conceptual contributions of our work is the introduction of a linear program that provides a {\em lower bound} on the performance of a given policy (\Cref{prop:plplowerbound}). To the best of our knowledge, this is the first study to achieve such a result. Previous studies have primarily focused on approximating an {\em upper bound} defined by a linear program. Our lower bound arises from a novel inequality established in \Cref{prop:gammabound}. Our approach allows us establish a competitive ratio of $1/2$ for all instances with homogeneous departures, generalizing the main result of \citet{kessel2022stationary}.

Additional contributions from our work include a slightly tighter upper bound for the omniscient benchmark (\Cref{prop:offline-bound}) and a proof that the solution of our LP is within a factor of 2 of this bound (\Cref{prop:factor2}). We show how to combine this last result with a result from \citet{aouad2022dynamic} to get a $\frac{1}{4}(1-1/e)$ approximation of the {\em omniscient} benchmark for the bipartite case  (see \Cref{sec:app:aouadguarantee}). As far as we know, this is the best guarantee for this case and is a slight improvement on the $1/8$ from \citet{collina2020dynamic} and 0.155 from \citet{li2023fully}.

\section{Model}
\label{sec:model}
An instance of a \textit{Matching Problem with Abandoment} is defined by the tuple $\mathcal{I} = \{\mathcal{T},\lambda,\mu, r \}$, where  $\m{T}$ is a finite set of agent types. 

{\bf Agent Arrival and Abandonment.}
Agents of type $i \in \m{T}$ arrive according to a Poisson process with rate $\lambda_i$. Arrivals are independent across types. For any time $t \ge 0$ and $i \in \m{T}$, let $\mathcal{A}_i(t)$ be the set of agents of type $i$ that arrive before time $t$. Let $\mathcal{A}(t) = \bigcup_{i \in \m{T}}\mathcal{A}_i(t)$. For each agent $a \in \mathcal{A}(t)$ , we let $i_a$ indicate the type of the agent, and $t_a$ indicate the agent's arrival time. The assumption of Poisson arrivals implies that conditioned on $|\mathcal{A}_i(t)|$, the times $\{t_a : a \in \mathcal{A}_i(t)\}$ are iid uniform on $[0,t]$. 



Each agent $a \in \mathcal{A}_i(t)$ has a {\em maximum sojourn time} $s_a$: after waiting for $s_a$ periods, agent $a$ will depart unmatched. We assume that agents of type $i$ abandon the system at rate $\mu_i$. Formally, this means that the times $\{s_a\}_{a \in \m{A}_i(t)}$ are iid and exponentially distributed with mean $1/\mu_{i}$. We let $d_a = t_a + s_a$ be the time at which agent $a$ will depart unmatched.

{\bf Rewards.} 
We can conduct pairwise matches of agents. The reward earned from forming a match depends on the types of the agents involved. Say that a match has type $(i,j) \in \mathcal{T} \times \mathcal{T}$ if the earlier-arriving  agent in the match is of type $i$ and the later-arriving agent in the match is of type $j$. (Because arrivals follow a Poisson process, with probability one no two agents arrive at the exact same time.) Let $r_{ij} \in \mathbb{R}$ specify the reward from making an $(i,j)$ match. We generalize previous work by allowing $r_{ij}$ to be different from $r_{ji}$. 

\subsection{Optimal Omniscient Policy}
\label{sec:offlineoptimal}

For any time $t\ge 0$, construct a weighted graph $G(t) = (\mathcal{A}(t), \mathcal{E}(t), w_t)$ where each vertex $a \in \mathcal{A}(t)$ represents an agent who arrived before time $t$, agents are connected by an edge if they were present at the pool at the same time, and the weight of each edge is the reward that would be obtained by matching these agents: 
\begin{align}
    \mathcal{E}(t) &= \{ (a,b) |a,b \in \mathcal{A}(t), a\not= b, [t_a, d_a] \cap [t_b, d_b] \not = \emptyset  \}    \\
    w_t(a,b) &= r_{i_a i_b} \quad \forall (a,b) \in \mathcal{E}(t).
\end{align}

\begin{definition}
    Let $\m{H}(t) = \{(i_a, t_a,d_a)\}_{a \in \mathcal{A}(t)}$ be the history up to time $t$. A policy $\pi$ maps the history $\m{H}(t)$ to a matching of $G(t)$. We use $\pi(\m{H}(t))$ to denote the matching chosen by policy $\pi$ at time $t$. 
\end{definition}
It is worth noting that we do not impose any informational constraints to $\pi$, nor do we require it to be ``consistent'' across time. For example, if only two agents arrive before time $t$, then when we optimize for time $t$, $\pi$ might match these two agents. When we optimize for a longer horizon, $\pi$ may choose to make a different match with these agents. 


We use $V(\pi, \m{I}, t)$ to be the expected reward rate of policy $\pi$ in instance $\m{I}$ by time $t\ge 0$.
\[V(\pi, \m{I}, t) = \frac{1}{t} \E{\sum_{(a,b) \in \pi(\m{H}(t))}w_t(a,b)} \]
Here the expectation is taken over the random arrival and departure process (that is, the sets $\mathcal{A}_i(t)$ and times $\{t_a,d_a\}_{a \in \mathcal{A}(t)}$, which induce the graph $G(t)$).
For any time $t>0$ and any graph $G(t)$, the omniscient policy always returns a maximum weight matching of $G(t)$.

\begin{definition}
    For any $t > 0$ and instance $\mathcal{I}$ the value of the optimal omniscient policy is:
    \begin{equation}
        \offoptt = \max_{\pi} V(\pi, \m{I}, t). 
    \end{equation}
\end{definition}
For policy $\pi$, we let $X_{ij}^\pi(t)$ be a counting process that tracks the number of $(i,j)$ matches:
\begin{equation}
    X_{ij}^\pi(t) = |\{(a,b) \in \pi(\m{H}(t)) \ | \ i_a = i, i_b = j, t_a < t_b\}|.
\end{equation}

\subsection{Greedy Policies}




A greedy policy $\pi = \{\succ_j \}_{j \in \m{T}}$ is defined by an ordering $\succ_j$ over $\m{T} \cup \{\emptyset \}$ for each $j \in \m{T}$. Whenever there is a new arrival of a type $j$, this greedy policy scans the list, and makes the highest-priority possible match according to $\succ_j$. For example, if $1 \succ_j 2 \succ_j \emptyset \succ_j 3$, then an arriving type $j$ will try to match with a type $1$, then with a type $2$. If neither of these types are available, the agent will simply join the pool.  Note that for every match made by a greedy policy, one of the agents involved in the match just arrived.\footnote{If there are multiple agents of the same type that could be part of a match, we choose one arbitrarily. Given the Markovian property of the process and the non-anticipatory nature of greedy policies, how we choose this makes no difference.} 


We use $N_i^\pi(t)$ to represent the number of agents of type $i$ that are present (have arrived but not yet matched or abandoned) at time $t$. 
Under any greedy policy $\pi$, $N^\pi(t) = \{N_i^\pi(t)\}_{i \in \mathcal{T}}$ is a continuous time Markov chain and has a unique steady-state distribution. (To shorten exposition, we defer the proof of this fact to \Cref{appendix:steady-state-greedy-policies}). We associate a greedy policy $\pi$ with the following quantities:
\begin{align}
    \label{def:ppi}
    n_i^\pi & = \lim_{t \to \infty} \E{N_i^\pi(t)}     \\
    \label{def:xpi}
    x_{ij}^{\pi} & = \lim_{t \to \infty} \frac{1}{t}\E{X_{ij}^\pi(t)}.
\end{align}
That is, $n_i^\pi$ is the expected number of agents of type $i$ present in the system in steady state and $x_{ij}^{\pi}$ is the rate at which $(i,j)$ matches are performed in steady-state. We use $N^\pi$ to denote a random variable drawn from the steady-state distribution of the chain $N^\pi(t)$.

The value of a policy can be expressed using the steady-state rate of matches $x_{ij}^\pi$: 
    \begin{equation}
        V(\pi, \mathcal{I}) = \lim_{t \to \infty} V(\pi, \m{I}, t) =  \sum_{i,j \in \mathcal{T}} x_{ij}^\pi r_{ij}.
    \end{equation}




\section{Algorithm}
\label{sec:procedure}
\subsection{Novel family of linear programs}
\label{sec:plpsec}
We start by introducing the family of linear programs $\plpm$ that we use to identify greedy policies and, in the case of homogenous departures, lower bound the reward rate of the identified policy. For a set of matches $\m{M} \subseteq \m{T} \times \m{T}$ define
\begin{equation}
    T(\m{M},j) = \{i \in \m{T} \ | \ (i,j) \in \m{M}\}
\end{equation}
to be the types that could match with an arriving $j$ according to $\m{M}$. Also, for $S \subseteq \mathcal{T}$ define $\gamma_S$ as in \eqref{eq:gammadef}. 
Our linear program has variables  ${\bf n} = \{n_i\}_{i \in \m{T}}$, ${\bf x} = \{x_{ij}\}_{(i,j) \in \m{M}}$ and slack variables $\bm{\psi} = \{\psi_{Sj}\}_{j \in \m{T}, S \subseteq T(\m{M},j) }$ and is defined as follows:
\begin{equation}
    \plpm = \left\{\max \limits_{{\bf n},{\bf x}, \bm{\psi}} \sum_{(i,j)\in \m{M}}r_{ij}x_{ij} \ | \   ({\bf n},{\bf x}, \bm{\psi}) \text{ satisfies } \eqref{eq:balance-lp-glb}, \eqref{eq:match-rate-lp-glb},  \eqref{eq:nonnegative-lp-glb} \right \}
\end{equation}
\begin{align}
    n_i \mu_i + \sum_{j : \,\,  (i,j) \in \m{M} } x_{ij} + \sum_{j : \,\, (j,i) \in \m{M} } x_{ji} &= \lambda_i & \quad \forall i \in \mathcal{T} \label{eq:balance-lp-glb}\\
    \sum_{i \in S} x_{ij} + \psi_{Sj} &= \lambda_j \gamma_{S}\sum_{i\in S} n_i & \quad \forall j \in \mathcal{T},  S \subseteq T(\m{M}, j) \label{eq:match-rate-lp-glb}\\
    x_{ij},\psi_{Sj} &\ge 0 & \quad \forall (i,j) \in \mathcal{M}, j \in \m{T}, S \subseteq T(\m{M},j)  \label{eq:nonnegative-lp-glb}
\end{align}
  Constraints \eqref{eq:balance-lp-glb} are balance equations that say that the arrival rate of a type $i$ must equal its match rate plus its abandonment rate $n_i \mu_i$. To understand the constraints \eqref{eq:match-rate-lp-glb} it useful to recall \Cref{prop:gammabound}, which says that under certain conditions,  $\PP{\sum_{i \in S} N_i^\pi > 0} \ge \gamma_S \sum_{i \in S} n_i^\pi$. Therefore, we can think of $\gamma_{S}\sum_{i\in S} n_i $ as a pessimistic approximation of the probability that an agent of some type $i \in S$ is present. 

\subsection{Suitable solutions and their connection to greedy policies}

Having introduced our family of linear programs, we now describe how to map a solution of these linear programs to a greedy policy. We consider two ways to do this. \begin{enumerate}
\item Determine $\succ_j$ based on which of the constraints in \eqref{eq:match-rate-lp-glb} are tight. Intuitively, if the constraint corresponding to $S$ and $j$ is tight, then $\textup{LP}^{\textup{ALG}}$ is trying to make as many matches as possible between arriving type $j$ agents and waiting agents from the set $S$, so types in $S$ should appear at the ``top'' of $\succ_j$.
\item Determine $\succ_j$ based on the ``value'' of each type, as determined by the dual of $\textup{LP}^{\textup{ALG}}$. Let $v_i$ be the dual variable associated with the $i^{th}$ constraint in \eqref{eq:balance-lp-glb}. Construct $\succ_j$ by ordering agents $i$ according to their ``score'' $r_{ij} - v_i - v_j$, with the acceptable matches being those for which this score is positive.
\end{enumerate}
For both methods, sometimes the solution to $\plpm$ cannot be readily mapped to a greedy policy. For example, sometimes there is no $i$ such that the constraint corresponding to $S = \{i\}$ and $j$ is tight, so it is unclear which type should be ranked first according to $\succ_j$. Similarly, sometimes the dual values are such that two types $i$ have the exact same score, or the score of a given match is exactly equal to zero. In other words, these methods work only if the solution of $\plpm$ is ``suitable.'' In this section, we define what it means for a solution of $\plpm$ to be suitable, and present an algorithm to find a set $\m{M}$ such that the corresponding solution is suitable. We focus on the primal-based method, saving discussion of the dual-based approach for \Cref{sec:dual-values}.

We start with a definition and some observations about greedy policies.
\begin{definition}
Set $S$ is a {\bf prefix} of order $\succ$ if it contains the $k$-most preferred types according to $\succ$ for some $k$. That is, there is some $i\succ \emptyset$ such that $S = \{i\} \cup \{j \in \m{T} \ | \ j\succ i \}$. We let $\prefixes{\succ}$ denote the set of all prefixes of $\succ$.
\end{definition}

So, for example, if $3 \succ 1 \succ \emptyset$, then $\prefixes{\succ} = \{\{3\}, \{1, 3\}\}$. 

\begin{lemma}
    \label{lemma:balance:multiple}
    Let $\pi$ be a greedy policy with preferences $\{\succ_j \}_{j \in \m{T}}$, and let $N^\pi$ be drawn from the steady state distribution of the Markov chain induced by $\pi$. Then
    \begin{align}
        n_i^\pi \mu_i + \sum_{j \in \m{T}} x_{ij}^\pi + x_{ji}^\pi &= \lambda_i & \quad  \forall i \in \m{T} \label{eq:balancegreedy1}\\
        \sum_{i \in S} x_{ij} &\le  \lambda_j \PP{\sum_{i \in S} N_i^\pi > 0} & \quad \forall j \in \m{T}, S \subseteq \m{M} \label{eq:messy}.
    \end{align}
    Furthermore, \eqref{eq:messy} holds with equality if and only if $S\in \prefixes{\succ_j}$. \\ Finally, if \eqref{eq:messy} holds with equality, then $x_{ij} > 0$ for each $i \in S$.
\end{lemma}
Note that \eqref{eq:balancegreedy1} and \eqref{eq:messy} parallel  \eqref{eq:balance-lp-glb} and \eqref{eq:match-rate-lp-glb}. Inspired by this, given  a solution to $\textup{LP}^{\textup{ALG}}$, we will construct a greedy policy such that $\prefixes{\succ_j}$ correspond to the sets $S$ such that \eqref{eq:match-rate-lp-glb} is tight.   The final claim from \Cref{lemma:balance:multiple} inspires the following definition.


\begin{definition}
    We say that a basic feasible solution  $({\bf n}, {\bf x}, \bm{\psi})$ of $\plpm$ is {\bf suitable} if $\psi_{Sj} = 0$ implies that $x_{ij} > 0$ for all $i \in S$.
\end{definition}

For a short refresher on basic feasible solutions, we refer the reader to \Cref{sec:refresher}. 

The following result establishes that for any suitable solution of $\textup{LP}^{\textup{ALG}}$, the sets $S$ such that $\psi_{Sj} = 0$ can be interpreted as prefixes of an order $\succ_j$. 
\begin{lemma}
    \label{lemma:suitablestructure}
    Let $({\bf n}, {\bf x}, \bm{\psi})$ be a suitable solution. For every $j \in \m{T}$ let 
    \begin{equation}
        \m{S}^j =  \{S \ : \ \psi_{Sj} = 0 \} \label{eq:Sj}
    \end{equation}
    For each $k \in \{1,2,..., |\m{S}^j| \}$ there is exactly one set $S_k^j \in \m{S}^j$ of size $k$, and $S_{k-1}^j \subset S_k^j$.
\end{lemma}
This result allows us to connect a suitable solution to a greedy policy, as follows.

\begin{definition}
Given a suitable solution $({\bf n}, {\bf x}, \bm{\psi})$ of $\plpm$, define the corresponding greedy policy $\pi$ by orderings $\{\succ_j\}_{j \in \m{T}}$ such that for all $j \in \m{T}$ we have $\prefixes{\succ_j} = \m{S}^j$.
\end{definition}

The proof of \Cref{lemma:balance:multiple} can be found in \Cref{proof:lemma:balance:multiple}, and the proof of \Cref{lemma:suitablestructure} can be found in \Cref{proof:lemma:suitablestructure}.


\subsection{Finding suitable solutions}

The preceding section showed that if the solution to $\plpm$ is suitable, it can be associated with a greedy policy. However, the optimal solution of $\plpm$ need not be suitable. Our next task is finding a set $\m{M}$ such that the optimal solution of $\plpm$ is suitable. We do this using \Cref{algo:suitability}. It starts by choosing $\m{M} = \m{T} \times \m{T}$. So long as the resulting optimal solution is not suitable, it identifies a match $(i,j)$ and a set $S$ containing $i$ such that $\psi_{Sj} = 0$ and $x_{ij} = 0$, and removes the match $(i,j)$ from $\m{M}$.


\begin{lemma}
    \label{lemma:algfindssuitable}
    \Cref{algo:suitability} terminates and returns a set $\m{M}$ where the optimal solution of $\plpm$ is suitable.
\end{lemma}
\begin{proof}
    In each step inside the while loop of \Cref{algo:suitability} either a match is removed or the loop ends. Because the number of matches to remove is at most $\m{T}^2$, the algorithm eventually terminates. Furthermore, a solution that makes no matches is suitable by definition (since $\psi_{Sj} > 0$ for all $S$), so if \Cref{algo:suitability} returns $\m{M}_k = \emptyset$, it is indeed finding an $\m{M}$ that leads to a suitable solution.
\end{proof}

\begin{algorithm}[t]
	\caption{Suitability finder} 
	\begin{algorithmic}[1]
        \State {\bf Input:} $\m{I} = \{\lambda, \mu, r \}$.
        \State $\m{M}_0 \leftarrow \m{T} \times \m{T}$.
        \State $R_0 \leftarrow 0$.
        \State $k \leftarrow 0$.
        \State   $not\_suitable \leftarrow \ True$.
        \While {$not\_suitable$}
        \State $k \leftarrow k+1$
        \State $R_k, ({\bf n, {\bf x}, \bm{\psi}}) \leftarrow \ solve \ \plpm$. \Comment{returns an optimal value and basic feasible solution. }
        \State $not\_suitable, match\_to\_remove \leftarrow \ check\_suitability({\bf n, {\bf x}, \bm{\psi}})$. \Comment{returns False and an empty set if the solution is suitable and True and a match to remove otherwise.  }
        \State $\m{M}_k = \m{M}_{k-1} \setminus match\_to\_remove $.
        \EndWhile
        \State {\bf Return:} $\m{M}_k$
	\end{algorithmic} 
    \label{algo:suitability}
\end{algorithm}

\subsection{Value-based policies} 
\label{sec:dual-values}

In this section we present a succinct representation of the greedy policy associated with an optimal suitable solution. We start by introducing a way to define greedy policies based on a single ``value'' for each type.

\begin{definition}
    \label{def:value-based-policy}
    Let $\m{M} \subseteq \m{T} \times \m{T}$ be a set of matches and ${\bf v} \in \mathbb{R}^{\m{T}}$ be a vector of values. We say that a greedy policy $\pi$ with orders $\{\succ_j \}_{j \in \m{T}}$ is {\bf consistent} with matches  $\m{M}$ and values ${\bf v}$ if for all $i, i', j \in \m{T}$, 
    \begin{itemize}
        \item if $i \succ_j \emptyset$ then $(i,j) \in \m{M}$
        \item if $r_{ij} - v_i - v_j > 0$ then $i \succ_j \emptyset$. 
        \item if $r_{ij} - v_i - v_j < 0$ then  $\emptyset \succ_j i $. 
        \item if $r_{ij} - v_i - v_j > r_{i'j} - v_{i'} - v_j \ge 0$ then $i \succ_j i'$  
    \end{itemize}
\end{definition}
For a set of feasible matches $\m{M}$ if we define the ``score'' of an $(i,j) \in \m{M}$ match to be $r_{ij} - v_i - v_j$, this says that we prioritize matches with higher scores, and do not form matches with negative scores.

It is worth noting that in some cases, multiple greedy policies could be consistent with a vector ${\bf v}$. For example, suppose you have a case were $r_{ij} - v_i - v_j = r_{i'j} - v_{i'} - v_j$ (that is, $(i,j)$ matches and $(i', j)$ matches have the same score) then both $i \succ_j i'$ and $i' \succ_j i$ are valid orders. Also, if the score $r_{ij} - v_i - v_j = 0$, then we have freedom to decide whether the policy considers $(i,j)$ matches to be acceptable. 

In the previous section we introduced an algorithm to find a greedy policy using suitable solutions of $\plpm$. We now show that the dual of $\plpm$ can be used to find values ${\bf v}$ that are consistent with this policy. The dual is as follows.



\begin{equation} \label{eq:dlpm}
        \dlpm = \left\{\min \limits_{{\bf v}, {\bf z}} \sum_{i \in \mathcal{T}} \lambda_i v_i \ |\ ({\bf v}, {\bf z}) \text{ satisfies }  \eqref{eq:dual-match-rate-cplt2}, \eqref{eq:dual-balance-cplt2}, \eqref{eq:nonnegative-dual-cplt2} \right\}.
\end{equation}

\begin{align}
    v_i + v_j + \sum_{  } z_{Sj}{\bf 1}(i \in S) &\ge r_{ij} \quad \forall (i,j) \in \mathcal{M} \label{eq:dual-match-rate-cplt2}\\
    v_i \mu_i &= \sum_{j \in \mathcal{T}}\sum_{S \subseteq T(\m{M}, j)}  \lambda_j \gamma_S z_{Sj} \quad \forall i \in \mathcal{T} \label{eq:dual-balance-cplt2}\\
    z_{Sj} & \ge 0 \quad \forall j \in \mathcal{T},  S \subseteq T(\m{M}, j) \label{eq:nonnegative-dual-cplt2}
\end{align}

\begin{lemma}
    \label{lemma:value-based}
    Let $\m{M}$ be a set of matches and $({\bf n}, {\bf x}, \bm{\psi})$ be an optimal suitable solution of $\plpm$ with associated greedy policy $\pi$. Also, let $({\bf v}, {\bf z})$ be an optimal solution of $\dlpm$. Then, the greedy policy $\pi$ is consistent with matches $\m{M}$ and values ${\bf v}$.
\end{lemma}
\begin{proof}
    For the first bullet in \Cref{def:value-based-policy} note that $r_{ij} - v_i - v_j > 0$ implies that there is at least one $S \subseteq T(\m{M}, j) $ where $i \in S$ and such that $z_{Sj} >0$ given that $({\bf v}, {\bf z})$ is feasible. By complementary slackness this implies that $\psi_{Sj} = 0$ and because the solution is suitable we have that $x_{ij} > 0$. Therefore, $i \succ_j \emptyset$.

    The second bullet in \Cref{def:value-based-policy} follows immediately from complementary slackness: 
    \begin{align}
        v_i + v_j + \sum_{ S \subseteq T(\m{M}, j)} z_{Sj}{\bf 1}(i \in S)-r_{ij} &\ge v_i + v_j > 0
    \end{align} 
    which immediately implies $x_{ij} = 0 $ and therefore $\emptyset \succ_j i$.

    Finally, we show the third bullet in \Cref{def:value-based-policy}. Fix a $j$ and take two types $i,i' \in \m{T}$ such that $r_{ij} - v_i - v_j > r_{i'j} - v_{i'} - v_j$. If $x_{i'j} = 0$ then $\emptyset \succ_j i'$ and as we showed in the first bullet $i \succ_j \emptyset$ and therefore by transitivity $i \succ_j i'$. We move to the more interesing case where $x_{i'j} >0$. By complementary slackness we have that
    \begin{align}
        v_i + v_j + \sum_{ S \subseteq T(\m{M}, j)} z_{Sj}{\bf 1}(i \in S) &= r_{ij}\\
        v_{i'} + v_j + \sum_{ S \subseteq T(\m{M}, j)} z_{Sj}{\bf 1}(i' \in S) &= r_{i'j}.
    \end{align} 
    Substracting the two equations and rearrenging gives us
    \begin{align}
        r_{ij} -v_i - v_j - \sum_{ S \subseteq T(\m{M}, j)} z_{Sj}{\bf 1}(i \in S) = r_{i'j} - v_{i'} - v_{j} - \sum_{ S \subseteq T(\m{M}, j)} z_{Sj}{\bf 1}(i' \in S).
    \end{align}
    By assumption we have that $r_{ij} - v_i - v_j > r_{i'j} - v_{i'} - v_{j}$, which implies that
    \begin{align}
        \sum_{ S \subseteq T(\m{M}, j)} z_{Sj}{\bf 1}(i \in S) > \sum_{ S \subseteq T(\m{M}, j)} z_{Sj}{\bf 1}(i' \in S).
    \end{align}
    Given the nested structure of $\m{S}^j$ shown in \Cref{lemma:suitablestructure}, this implies that $\{S \in \m{S}^j : i' \in S \} \subset \{S \in \m{S}^j : i \in S \}$, which in turn implies that $i \succ_j i'$.
\end{proof}

As a final remark we note that if the optimal dual solution is nondegenerate, then the values ${\bf v}$ identify a unique greedy policy. However, if the optimal dual solution is degenerate, then there could be multiple greedy policies that follow the same values ${\bf v}$. The technique presented in \cite{wolfe1963technique} of adding small perturbations of the $r_{ij}$ values can be used to break the degeneracy and identify a greedy policy.

\section{Results}

\label{sec:results}
The previous sections described the methods used to identify greedy policies. We succintly describe the entire procedure in the following definition.
\begin{definition}
    For an instance $\m{I}$, $\alg$ is the greedy policy corresponding to the optimal suitable solution of $\plpm$, where $\m{M}$ is the result of running \Cref{algo:suitability}.
\end{definition}
In this section we present the intermediate results that jointly imply \Cref{thm:emu:result}: $\alg$ achieves at least half of the omniscient benchmark if departures are homogeneous.


\subsection{Lower bound for $\alg$}
We now show that if we have homogenous departures and the optimal solution of $\plpm$ is suitable, then it provides a lower bound on the value of the corresponding greedy policy.

\begin{proposition}
    \label{prop:plplowerbound}
    Let $\m{I}$ be an instance and $\m{M}$ be a set of matches. Let $({\bf n}^*, {\bf x}^*, \bm{\psi}^*)$ be an optimal and suitable solution of $\plpm$ and $\pi$ be its corresponding greedy policy with associated $n^\pi$. If for all $j \in \m{T}$ and all $S \subseteq \m{T}$ such that $\psi^*_{Sj} = 0$ we have 
    \begin{equation} \mathbb{P}(\sum_{i \in S} N_i^\pi > 0) \ge \gamma_S \sum_{i \in S} n_i^\pi, \label{eq:alpha-gamma} \end{equation}
    then $V(\pi, \m{I}) \ge \plpm$.
\end{proposition}

Recall that $\psi^*_{Sj} = 0$ if and only if $S$ is a prefix of the set of acceptable matches for $j$. If we have homogeneous departures (\Cref{def:homogeneous}), then \Cref{prop:gammabound} states that \eqref{eq:alpha-gamma} holds. However, even without this assumption, \eqref{eq:alpha-gamma} will sometimes be satisfied. (For example, this will be the case if each type has at most one acceptable match.) Because none of the remaining steps of our proof require any assumptions on the instance $\m{I}$, in these instances, if $\m{M}$ is the subset of matches identified by \Cref{algo:suitability}, then the reward rate of the resulting greedy algorithm is guaranteed to be at least half of the omniscient benchmark.


Although the conditions of \Cref{prop:plplowerbound} are not always satisfied, we have yet to find an instance $\m{I}$ for which its conclusion fails to hold. In \Cref{sec:numerical}, we present numerical data from randomly generated instances which do not have homogeneous departure rates. In all of these instances, $V(\alg, \m{I}) \ge \plpm$.

\begin{SCfigure}
\includegraphics[width=.5 \textwidth]{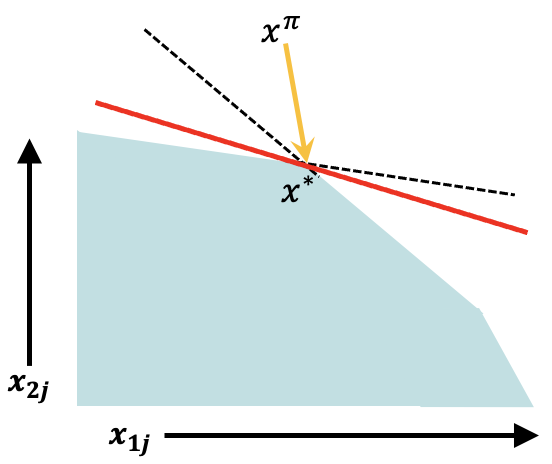}
\caption{Visualization of \Cref{glb:boundedbypolicy}. The blue region represents the polytope of feasible solutions to $\plpm$. The point $x^*$ is a vertex of this polytope. The red line is determined by the rewards $r$, and represents the set of points which have the same objective value as $x^*$. Although $x^\pi$ may not component-wise dominate $x^*$, \Cref{glb:boundedbypolicy} shows that the ``arrow'' from $x^\pi$ to $x^*$ ``pierces'' the feasible region, which implies that $x^\pi$ has a higher objective value than $x^*$ (lies above the red line).} \label{fig:piercing}
\end{SCfigure}

To prove \Cref{prop:plplowerbound}, we compare the optimal $\bx^*$ to the match rates $\bx^\pi$ induced by the corresponding greedy policy. The point $\bx^\pi$ will not be a feasible point of $\plpm$. Although \Cref{prop:gammabound} states that $\gamma_{S} \sum_{i \in S} n_i$ is an {\em underestimate} of the probability that an agent of some type $i \in S$ will be available, it is not the case that $\bx^*$ is component-wise smaller than (dominated by) $\bx^\pi$. Instead, our proof of \Cref{prop:plplowerbound} relies on the following Lemma, which in effect says that the ``arrow'' from $\bx^\pi$ to $\bx^*$ ``pierces'' the feasible region of $\plpm$. This is visualized in \Cref{fig:piercing}, and immediately implies that for any reward matrix $r$ such that $\bx^*$ is optimal, we have $r \circ \bx^\pi \ge r \circ \bx^*$.

\begin{lemma}
    \label{glb:boundedbypolicy}
    Let $({\bf n}, {\bf x}, \bm{\psi})$ be a suitable solution of $\plpm$, and $\pi$ its corresponding greedy policy with associated  $({\bf n}^\pi, {\bf x}^\pi)$. For $j \in \m{T}$, let $\m{S}^j = \{S \subseteq\m{T} : \psi_{Sj} = 0\}$, and let $\m{S} = \bigcup_{j \in \m{T}} \m{S}^j $.  If \eqref{eq:alpha-gamma} holds for all $S \in \m{S}$, then there is a $\beta \in (0,1)$ and a feasible point $(\hat{{\bf n}}, \hat{{\bf x}}, \hat{\bm{\psi}})$ of $\plpm$ such that $\beta {\bf x}^\pi + (1-\beta) \hat{{\bf x}} = {\bf x}$.
\end{lemma}

\begin{proof}

    Given any $\beta \in (0,1)$ define $(\hat{{\bf n}}, \hat{{\bf x}}, \hat{\bm{\psi}})$ as follows:
    \begin{align}
        \hat{x}_{ij} &= \frac{x_{ij} - \beta x_{ij}^\pi}{1-\beta} & \quad \forall (i,j) \in \m{M}, \\
        \hat{n}_i &= \frac{\lambda_i - \sum_{j \in \m{T}: (i,j) \in \m{M}}\hat{x}_{ij} - \sum_{j\in\m{T} : (j,i) \in \m{M}}\hat{x}_{ji} }{\mu_i} & \quad \forall i \in \m{T},\label{eq:hatni}\\
        \hat{\psi}_{Sj} &= \lambda_j \gamma_S \sum_{i \in S} \hat{n}_i - \sum_{i \in S} \hat{x}_{ij} & \quad \forall j \in \m{T}, S \subseteq T(M,j). \label{eq:hatpsiSj}
    \end{align}
    We will show that $(\hat{{\bf n}}, \hat{{\bf x}}, \hat{\bm{\psi}})$ is feasible for all sufficiently small $\beta$. By construction we can see that constraints \eqref{eq:balance-lp-glb} and \eqref{eq:match-rate-lp-glb} of $\plpm$ hold, so it remains to show that \eqref{eq:nonnegative-lp-glb} holds: $\hat{x}_{ij}$ and $\hat{\psi}_{Sj}$ are non-negative.
    
     If $x_{ij} = 0$, then because $({\bf n}, {\bf x}, \bm{\psi})$  is suitable, we know for any $S$ that contains $i$, $\psi_{Sj} > 0$, and therefore in the greedy policy $\pi$, $i$ is not acceptable to $j$. Therefore $x_{ij}^\pi = 0$ and $\hat{x}_{ij} = 0$. Meanwhile, if $x_{ij} > 0$, then $\hat{x}_{ij} > 0$ for all sufficiently small $\beta$.

    It remains to show that $\hat{\psi}_{Sj} \ge 0$. We start by rewriting \eqref{eq:hatni} and \eqref{eq:hatpsiSj} as follows.
      \begin{align}
        \hat{n}_i & = \frac{n_i - \beta n_i^\pi}{1-\beta}.\label{eq:equalunmatchedrate} \\
        \hat{\psi}_{Sj} & = \frac{\psi_{Sj} - \beta(\lambda_j\gamma_S \sum_{i \in S} n_i^\pi - \sum_{i \in S} x_{ij}^\pi)}{1-\beta}. \label{eq:hatpsiSj-rewritten}
    \end{align}
        Equation \eqref{eq:equalunmatchedrate} follows because \eqref{eq:balance-lp-glb} holds for $({\bf n}, {\bf x}, \bm{\psi})$ and $(\hat{{\bf n}}, \hat{{\bf x}}, \hat{\bm{\psi}})$ and \eqref{eq:balancegreedy1} holds for $({\bf n}^\pi, {\bf x}^\pi)$ by \Cref{lemma:balance:multiple}. Meanwhile,  \eqref{eq:hatpsiSj-rewritten} follows from the definition of $\hat{x}_{ij}$ and from \eqref{eq:equalunmatchedrate}.
    
Examining \eqref{eq:hatpsiSj-rewritten}, we see that if $\psi_{Sj} > 0$ then for all sufficiently small $\beta$, $\hat{\psi}_{Sj} > 0$. If $\psi_{Sj} = 0$ then by \eqref{eq:messy} in \Cref{lemma:balance:multiple} we have that $\sum_{i \in S} x_{ij}^\pi = \lambda_j\PP{\sum_{i \in S} N_i^\pi > 0}$, from which the assumption that \eqref{eq:alpha-gamma} holds implies that, \[ \lambda_j \gamma_S\sum_{i \in S} n_i^\pi -\sum_{i \in S} x_{ij}^\pi \le 0.\] Thus, $\hat{\psi}_{Sj} \ge 0$ in these cases as well. We conclude that for sufficiently small $\beta$, $(\hat{{\bf n}}, \hat{{\bf x}}, \hat{\bm{\psi}})$ is a feasible solution of $\plpm$.
\end{proof}

\begin{proof}[Proof of \Cref{prop:plplowerbound}]
    Let $({\bf n}^\pi, {\bf x}^\pi)$ be associated with the greedy policy $\pi$. By assumption, the conditions of \Cref{glb:boundedbypolicy} are satisfied and therefore, there exists a feasible solution $(\hat{{\bf n}}, \hat{{\bf x}}, \hat{\bm{\psi}})$ of $\plpm$ and a $\beta \in (0,1)$ such that  $\beta {\bf x}^\pi + (1-\beta) \hat{{\bf x}} = {\bf x}$. Because $(\hat{{\bf n}}, \hat{{\bf x}}, \hat{\bm{\psi}})$ is feasible we must have that $r \circ \hat{{\bf x}} \le r \circ {\bf x}$. Because $\beta \in (0,1)$ and $r \circ (\beta {\bf x}^\pi + (1-\beta) \hat{{\bf x}}) = r \circ {\bf x}$ it follows that $r \circ {\bf x}^\pi \ge r \circ {\bf x}$. This completes the proof, as $r \circ \bx^\pi = V(\pi, \m{I})$ and $r \circ \bx^*= \plpm$.
\end{proof}



\subsection{New upper bound for any policy}
Our second step is to present two linear programs, both of which upper bound the value of the omniscient benchmark. 

    \begin{equation}
        \offrb = \max_{{\bf x}}  \left\{ \sum_{i,j \in \mathcal{T}} r_{ij} x_{ij} | {\bf x} \text{ satisfies } \eqref{eq:offlinematchrate2}, \eqref{eq:posrate2}  \right\}.
    \end{equation}
    \begin{align}
         \sum_{i \in S} x_{ij} + \sum_{i \in S'} x_{ji} &\le \lambda_j \left(1 - \frac{\mu_j}{\mu_j + \sum_{i \in S'} \lambda_j} e^{-\sum_{i \in S}\lambda_i/\mu_i}\right) & \quad \forall j \in \mathcal{T}, S,S'\subseteq \mathcal{T} \label{eq:offlinematchrate2}\\
        x_{ij} & \ge 0 &  \quad \forall i,j \in \mathcal{T} \label{eq:posrate2}
    \end{align}
    
    \begin{equation}
    \offrbrel =  \left\{\max_{{\bf n}, {\bf x}}\sum_{i,j \in \mathcal{T}}r_{ij}x_{ij} \ | \ \text{satisfies } \eqref{eq:balance-offline-relax2}, \eqref{eq:match-rate-offline-relax2}, \eqref{eq:nonnegative-offline-relax2} \right\}
\end{equation}
\begin{align}
    \sum_{i \in \mathcal{T}}x_{ij} +   \sum_{i \in \mathcal{T}}x_{ji}&\le \lambda_j & \quad \forall j \in \mathcal{T} \label{eq:balance-offline-relax2}\\
    \sum_{i \in S} x_{ij} &\le \lambda_j \left(1-e^{- \sum_{i \in S} \frac{\lambda_i}{\mu_i}}  \right) & \quad \forall j \in \mathcal{T}, S \subseteq \mathcal{T} \label{eq:match-rate-offline-relax2}\\
    x_{ij} &\ge 0 & \quad \forall i,j \in \mathcal{T} \label{eq:nonnegative-offline-relax2}
\end{align}

\begin{restatable}{proposition}{offlinebound}
    \label{prop:offline-bound}
    For all instances $\mathcal{I}$ we have $\offopt \le \offrb \le \offrbrel$.
\end{restatable}


In \Cref{sec:1/2guarantee}, we use $\offrbrel$ to establish a competitive ratio of $1/2$. The idea behind \eqref{eq:match-rate-offline-relax2} is that an arriving agent of type $j$ can match to a waiting agent of type $i \in S$ only if such an agent is present in the system, and the probability of this event is at most $1-e^{- \sum_{i \in S} \frac{\lambda_i}{\mu_i}}$.

The linear program $\offrbrel$ provides a tighter bound than those used by  \citet{collina2020dynamic} and \citet{kessel2022stationary}, and is more general than the linear program used by \citet{amanihamedani2024improved}. Specifically, \citet{collina2020dynamic} consider only singleton $S$ in \eqref{eq:match-rate-offline-relax2}, and use $\lambda_j \lambda_i/\mu_i$ as the upper bound instead of $\lambda_j (1-e^{-\lambda_i/\mu_i})$. \citet{kessel2022stationary} also consider only singleton $S$, but use the bound $\lambda_j (1-e^{-\lambda_i/\mu_i})$. The linear program from \citet{amanihamedani2024improved} considers all subsets $S$, but assumes that the graph of possible matches is bipartite, and that one side must be matched immediately. Our linear program $\offrbrel$ is equivalent to theirs for this special case,\footnote{Because their primary focus is to benchmark against the optimal {\em online} policy, their linear program includes an additional constraint, which they drop when comparing to the optimal offline policy. We compare to their linear program without this constraint.} but also applies to non-bipartite graphs in which each type has an abandonment rate $\mu_i < \infty$.

We do not directly use $\offrb$, but present it here because it provides an even tighter bound than those used in past work, and incorporates a new idea. Specifically, whereas \eqref{eq:match-rate-offline-relax2} in $\offrbrel$ considers only the possibility that an agent of type $j$ matches to a previous arrival, \eqref{eq:offlinematchrate2} considers the possibility that an arriving agent may match with either a previous arrival or a future arrival. Matches to future arrivals are incorporated using the set $S'$, and taking $S' = \emptyset$ in \eqref{eq:offlinematchrate2} recovers the constraints \eqref{eq:match-rate-offline-relax2}. Meanwhile, taking $S = S' = \mathcal{T}$ in \eqref{eq:offlinematchrate2} recovers a strengthening of \eqref{eq:balance-offline-relax2}, as the resulting right side is strictly less than $\lambda_j$. It follows that $\offrbrel$ is a relaxation of $\offrb$, and therefore $ \offrb \le \offrbrel$.

The complete proof of \Cref{prop:offline-bound} can be found in \Cref{proof:prop:offline-bound}. We present a proof sketch that captures the main idea behind this result. 

\begin{proof}[Proof sketch]
    Suppose agent $a$ of type $j$ arrives at time $t$ and would abandon at time $d_a$. This agent can only contribute towards the the sum on the left side of \eqref{eq:offlinematchrate2} if there is an agent of some type in $i \in S$ present at time $t$ (enabling an $(i,j)$ match), or if an agent of some type $i \in S'$ arrives between $t$ and $d_a$ (enabling a $(j,i)$ match). 


The number of agents of type $i$ with an arrival time before $t$ and a final departure time after $t$ can be shown to follow a Poisson distribution with mean $\frac{\lambda_i}{\mu_i} (1-e^{-t\mu_i}) \le \frac{\lambda_i}{\mu_i}$. It follows that the probability that there is {\em no} agent of any type $i \in S$ available to match with $a$ at time $t$ is at least $e^{-\sum_{i \in S} \lambda_i/\mu_i}$.



Similarly, because the maximum sojourn time of $a$ is exponentially distributed with rate $\mu_j$ and the time until the next arrival of some agent $i\in S'$ is exponentially distributed with rate $\sum_{i\in S'} \lambda_i$, the probability that $a$ departs before the next arrival from $S'$ is $\frac{\mu_j}{\mu_j + \sum_{i \in S'} \lambda_i}$.

Noting that the events ``there is an agent of some type $i \in S$ present at time $t$'' and ``an agent of some type $i \in S'$ arrives after $t$ and before $d_a$'' are independent, we conclude that the probability that $a$ can contribute to the the sum on the left side of \eqref{eq:offlinematchrate2} is at most $(1 - \frac{\mu_j}{\mu_j + \sum_{i \in S'} \lambda_i}e^{-\sum_{i \in S} \lambda_i/\mu_i})$. Multiplying by the arrival rate of type $j$ agents yields \eqref{eq:offlinematchrate2}.
\end{proof}


\subsection{Approximating $\offrbrel$} \label{sec:1/2guarantee}
The final ingredient of our main result is showing that $\plpm$ is always at least half of $\offrbrel$.


\begin{proposition}
For any instance $\mathcal{I}$ let $\m{M}$ be the set of matches identified by \Cref{algo:suitability}. Then $\offrbrel \le 2 \cdot \plpm$ \label{prop:factor2}
\end{proposition}

To prove this result, we claim that
\[ \offrbrel \le 2 \cdot \plpt \le 2  \cdot \plpm.\]
\Cref{lemma:rb2plpt} establishes the first inequality and \Cref{lemma:algisincreasing} establishes the second, completing the proof of \Cref{prop:factor2}.

\begin{lemma}
    \label{lemma:rb2plpt}
    For any instance $\m{I}$, $\offrbrel \le 2 \cdot \plpt$.
\end{lemma}

\begin{proof}
The proof follows from strong duality. We introduce the duals of $\plpt$ and $\offrbrel$.
\begin{equation} \label{eq:dlpt}
    \dlplong = \left\{\min \limits_{{\bf v}, {\bf z}} \sum_{i \in \mathcal{T}} \lambda_i v_i \ |\ ({\bf v}, {\bf z}) \text{ satisfies }  \eqref{eq:dual-match-rate-cplt}, \eqref{eq:dual-balance-cplt}, \eqref{eq:nonnegative-dual-cplt} \right\}.
\end{equation}
\vspace{-.2 in}
\begin{align}
    v_i + v_j + \sum_{S \subseteq \m{T}} z_{Sj}{\bf 1}(i \in S) &\ge r_{ij} \quad \forall i,j \in \mathcal{T} \label{eq:dual-match-rate-cplt}\\
    v_i \mu_i &= \sum_{j \in \mathcal{T}} \sum_{S  \subseteq \m{T}} \lambda_j \gamma_S z_{Sj}{\bf 1}(i \in S) \quad \forall i \in \mathcal{T} \label{eq:dual-balance-cplt}\\
    z_{Sj} & \ge 0 \quad \forall S \subseteq \mathcal{T}, j \in \mathcal{T} \label{eq:nonnegative-dual-cplt}
\end{align}
    \begin{equation}
        \offrdrel =  \left \{\min_{{\bf v}, {\bf z}}\sum_{i \in \mathcal{T}} v_i \lambda_i + \sum_{j \in \mathcal{T}} \sum_{S\subseteq \mathcal{T}} z_{Sj}\lambda_j\left(1-e^{-\sum_{i \in S}\lambda_i /\mu_i } \right) \ | \ ({\bf v}, {\bf z}) \text{ satisfies } \eqref{eq:off-dual-matchrate2}, \eqref{eq:off-dual-nonnegative2} \right\}
    \end{equation}
    \begin{align}
        v_i + v_j + \sum_{S \subseteq \m{T}} z_{Sj} {\bf 1}(i \in S)&\ge r_{ij} \quad \forall i,j \in\mathcal{T} \label{eq:off-dual-matchrate2}\\
        v_i, z_{Sj} &\ge 0 \quad \forall i,j\in\mathcal{T}, S \subseteq \mathcal{T} \label{eq:off-dual-nonnegative2}.
    \end{align}

    We can multiply both sides of \eqref{eq:dual-balance-cplt} by $\lambda_i/\mu_i$, add the resulting constraints across all types $i \in \m{T}$ and use the definition of $\gamma_S$ in \eqref{eq:gammadef} to get that for any feasible solution $({\bf v}, {\bf z})$ to $ \dlplong$,
    \begin{align}
        \sum_{i \in \mathcal{T}}\lambda_i v_i  & \ge \sum_{i \in \mathcal{T}}\sum_{j \in \mathcal{T}} \sum_{S \subseteq \m{T}} \lambda_j  z_{Sj} {\bf 1}(i \in S)\left(1 -e^{-\sum_{k \in S} \lambda_k/\mu_k}\right)\frac{\lambda_i/\mu_i}{\sum_{k \in S} \lambda_k/\mu_k} \nonumber \\
         & = \sum_{j \in \mathcal{T}} \sum_{S \subseteq \mathcal{T}} \sum_{i \in S} \lambda_j  z_{Sj} \left(1-e^{-\sum_{k \in S} \frac{\lambda_k}{\mu_k}} \right) \frac{\lambda_i/ \mu_i}{ \sum_{k \in S} \lambda_k/\mu_k} \nonumber \\
 & =  \sum_{j \in \mathcal{T}} \sum_{S \subseteq \mathcal{T}} \lambda_j  z_{Sj} \left(1-e^{-\sum_{k \in S} \frac{\lambda_k}{\mu_k}} \right). \label{lemma:ineq:297}
    \end{align}
    We use this result as follows. Take an optimal solution $({\bf v^*}, {\bf z^*}) \in \argmin \dlplong$ and note that it is a feasible solution of $\offrdrel$. We then have that:
    \begin{align}
        \offrbrel & = \offrdrel \nonumber \\
        &\le \sum_{i \in \mathcal{T}} \lambda_i v^*_i + \sum_{j \in \mathcal{T}} \sum_{S\subseteq \mathcal{T}} z^*_{Sj}\lambda_j\left(1-e^{-\sum_{i \in S}\lambda_i /\mu_i } \right) \nonumber\\
        &\le 2\sum_{i \in \mathcal{T}} \lambda_i v_i^* \nonumber\\
        & = 2\cdot \dlplong  \nonumber \\
        & = 2\cdot \plpt 
    \end{align}
    where the first and last equalities come from strong duality, the first inequality comes from the fact that $({\bf v^*}, {\bf z^*})$ is a feasible solution of $\offrdrel$ and the second inequality comes from \eqref{lemma:ineq:297}.
\end{proof}

\begin{lemma}
    \label{lemma:algisincreasing}
    The optimal values $R_k$ defined in \Cref{algo:suitability} form a nondecreasing sequence.
\end{lemma}
\begin{proof}
    Consider iteration $k$ of the while loop with set of matches $\m{M}_k$. The algorithm finds an optimal solution $({\bf n}, {\bf x}, \bm{\psi})$. If the solution is not suitable then it returns $i, j \in \m{T}$ such that $x_{ij} = 0$ and there exists some $S \subseteq \m{T}$ containing $i$ such that $\psi_{Sj} = 0$. In this case, we have $\m{M}_{k+1} = \m{M}_k \setminus \{(i,j)\}$.
    
Note that because $x_{ij} = 0$, we can construct a feasible solution to $\plp(\m{I},\m{M}_{k+1})$ from $({\bf n}, {\bf x}, \bm{\psi})$ by simply removing the variables $x_{ij}$ and $\psi_{Sj}$ for $S$ such that $i \in S$, and keeping the value of all other variables the same. Because $x_{ij} = 0$, this solution has an objective value of $R_k$. Therefore the value of the optimal solution to $\plp(\m{I},\m{M}_{k+1})$ must be $R_{k+1} \ge R_k$.
\end{proof}

\section{Numerical Experiments}
\label{sec:numerical}
\begin{figure}[H]
    \centering
    \includegraphics[width=0.8\textwidth]{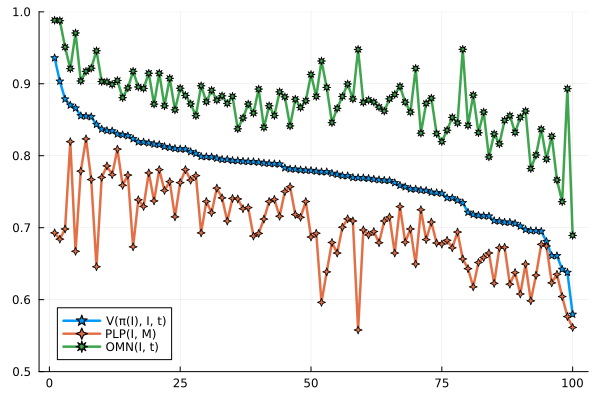}
    \caption{Comparison between $\plpm$, $V(\tilde{\pi}(\m{I}), \m{I}, t)$, $\offoptt$ and $\offrb$ for $100$ simulations with $|\m{T}| = 3$ and $t=100000$. }
    \label{fig:threetypes}
\end{figure}

\begin{figure}[H]
    \centering
    \includegraphics[width=0.8\textwidth]{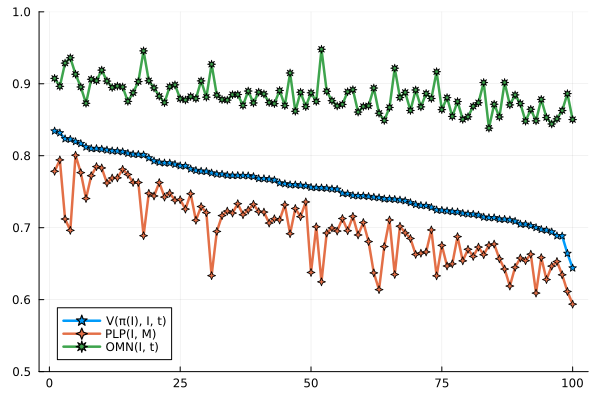}
    \caption{Comparison between $\plpm$, $V(\tilde{\pi}(\m{I}), \m{I}, t)$, $\offoptt$ and $\offrb$ for $100$ simulations with $|\m{T}| = 6$ and $t=100000$.  }
    \label{fig:sixtypes}
\end{figure}

\begin{figure}[H]
    \centering
    \includegraphics[width=0.8\textwidth]{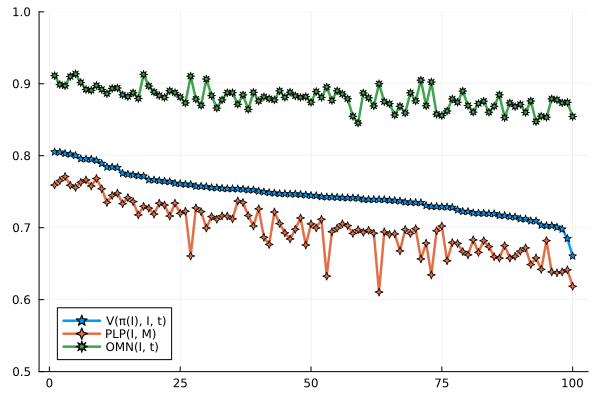}
    \caption{Comparison between $\plpm$, $V(\tilde{\pi}(\m{I}), \m{I}, t)$, $\offoptt$ and $\offrb$ for $100$ simulations with $|\m{T}| = 10$ and $t=100000$.  }
    \label{fig:tentypes}
\end{figure}

The objective of presenting numerical experiments is twofold: (i) it provides some evidence that $\plpm$ is a lower bound to the corresponding greedy policy even in instance that do not have homogenous departures and (ii) they give insight on how tight are the different approximations presented in this work. In Figures \ref{fig:threetypes}, \ref{fig:sixtypes} and \ref{fig:tentypes} we present the results of $100$ simulations. Each figure presents the results for a different number of types. In each simulation we sample $\bar{\lambda_i} \sim Unif(0,1)$ and define $\lambda_i = \frac{\bar{\lambda_i}}{\sum_{j \in \m{T}} \bar{\lambda_j}}$,  $\mu_i \sim Unif(0.01, 4)$ 
and $r_{ij} \sim 6 \cdot Unif(0,1)^2$ 
We normalize the values of $\plpm$, $V(\tilde{\pi}, \m{I}, t)$ and $\offoptt$ by dividing them by $\offrb$. We sort the simulations by $V(\tilde{\pi}, \m{I}, t)/\offrb$. 

These figures show that $\plpm$ is a lower bound to the corresponding greedy policy, at least in the cases simulated. This gives some evidence that $\plpm$ might give a lower bound even in instances with non-homogenous departures. It is worth pointing out that in the simulations studied, the condition required in \Cref{glb:boundedbypolicy}, (namely that for any $S\subseteq \m{T}$ where $\psi_{Sj} =0$  for some $j$ we have that $\PP{\sum_{i \in S} N_i^\pi > 0} \ge \gamma_S \sum_{i \in S} n_i^\pi$) holds. This is not the case for any Markovian policy. In \Cref{sec:app:alphasmaller} we present an example of a Markovian policy for which this condition does not hold. We are not aware of any counter-examples when the policy $\pi$ is greedy. This is worth exploring further because if $\PP{\sum_{i \in S} N_i^\pi > 0} \ge \gamma_S \sum_{i \in S} n_i^\pi$ for any greedy policy $\pi$ and set $S$ then we could expand our results to general instances (as this is the only part of our proof where the homogenous departure assumption is needed).

It is also interesting to note that, especially in Figures \ref{fig:sixtypes} and \ref{fig:tentypes}, $\plpm$ is a pretty tight lower bound to the greedy policy. Similarly,  $\offrb$ is not far from $\offoptt$ specially in cases where there are more types. 

\section{Discussion}


\label{sec:discussion}
In this work we presented an algorithm to find greedy policies that can guarantee at least half of the reward earned by the omniscient benchmark. While aspects of our approach parallel prior work, our key innovation lies in using a linear program ($\textup{LP}^\textup{ALG}$) that provides guaranteed {\em lower bounds} on policy performance. 

The most obvious limitation of our work is that we require strong assumptions on agents' departure rates. 
We are uncertain whether these assumptions are truly necessary, and hope to extend our result for cases with general $\mu_i$. In particular, we present the following conjecture. 
\begin{conjecture}  \label{con:greedy}
    For any instance $\mathcal{I}$ our proposed greedy policy $\alg$  achieves at least half of the omniscient reward rate:
    \[V(\alg, \mathcal{I})\ge \frac{1}{2} \offopt.\]
\end{conjecture}
Because \Cref{thm:no-better-half} shows that a guarantee greater than $1/2$ is impossible for any online policy, this would imply that deterministic greedy policies suffice to get an optimal competitive ratio. The current limitation in extending of our approach lies in the fact that we don't know if $\PP{\sum_{i \in S} N_i^\pi > 0} \ge \gamma_S \sum_{i \in S} \mathbb{E}[N_i^\pi]$ holds for greedy policies. We know that this inequality does not hold for general Markovian policies, as shown in the example in \Cref{sec:app:alphasmaller}.


\vspace{-.3 in}
\bibliography{refs}

\newpage
\appendix
\section{Refresher on basic feasible solutions}
\label{sec:refresher}
In this section we present a refresher on basic feasible solutions. We start by defining the standard linear program. Let $A \in \mathbb{R}^{m\times n}$ where $m$ is the number of constraints and $n$ is the number of variables and we assume that $n > m$ and that all $m$ rows of $A$ are linearly independent. Let $b \in \mathbb{R}^m$ be the right hand side of the constraints and $c \in \mathbb{R}^n$ be the cost vector. The standard linear program can be written as follows:

\begin{align}
    \max_{{\bf x}} \left\{ c^T {\bf x} \ | \ A {\bf x} =  b, \ {\bf x} \ge 0 \right\}
\end{align}

Let $\m{B} \subset \{1,2,...,n\}$ be a subset of $m$ indices. We use $A_{\m{B}}$ to be the square $m\times m$ submatrix of $A$ resulting on taking the columns of $A$ indexed by $\m{B}$. A solution ${\bf x}$ is a basic feasible solution of basis $\m{B}$ if $A_{\m{B}}$ is full rank, $x_{i} = 0$ for all $i \not \in \m{B}$, $A {\bf x} =  b$ and ${\bf x} \ge 0$. For a basis $\m{B}$ we use $x_{\m{B}}$ to denote the basic variables, that is the subset of variables that can be non-zero. In particular we can calculate the basic variables as follows: 
\begin{equation}
    x_{\m{B}} = A_{\m{B}}^{-1} b
\end{equation}

\section{Additional Proofs} \label{appendix:steady-state-greedy-policies}

All irreducible continuous time Markov chains on a finite state space have a unique stationary distribution. Because we have a countably infinite state space, we need to show that the chain $N^\pi$ is stable: it cannot ``drift upwards'' forever (as does, for example, a biased simple random walk on the integers). In our setting, this cannot occur because all of the abandonment rates $\mu_i$ are positive. Therefore, whenever the number of agents in the system becomes large, there is negative drift, ensuring that the system is stable. In this section, we establish this fact more formally. We make use of the following result.

\begin{theorem*}{\citep[Theorem 2.8]{prieto2016uniform}}
    Suppose that $N$ is an irreducible continuous time Markov chain with state space $\m{N}$ and transitions from state $n$ to $n'$ at rate $q_{n n'}$, and suppose that there exist a nonnegative function $V$ on $\m{N}$, a function $w: \m{N} \rightarrow [1,\infty)$, a finite set $C \subset \m{N}$ and constants $c > 0$ and $b \in \mathbb{R}$ such that for all $n \in \m{N}$,
    \begin{equation}
        \sum_{n' \in \m{N}} q_{nn'} (V(n')-V(n)) \le -cw(n) + b \mathbb{I}(n \in C). \label{eq:stability}
    \end{equation} 
    Then the Markov chain $N$ is ergodic.
\end{theorem*}

\begin{lemma}
    \label{lemma:stationaryctmc}
    Any greedy policy is Markovian. If $\pi$ is a Markovian policy, then $N^\pi(t)$ is ergodic and therefore has a unique stationary distribution.  
\end{lemma}

\begin{proof}
    We first show that for any greedy policy $\pi$, $N^{\pi}(t)$ is a continuous time Markov chain. This is because transitions will occur only when an agent arrives or abandons the system. Therefore, from state $n$, the time until the next transition follows an exponential distribution with mean $\sum_{i \in \mathcal{T}} (\lambda_i + n_i \mu_i)$. Furthermore, the state to which the system transitions depends only on the current state $n$, and not the history of the process.

    We now assume that $\pi$ is Markovian, and define $\mathcal{N}^\pi$ to be the set of states that are reachable from the empty state under policy $\pi$. (Note that $\m{N}^\pi$ does indeed depend on $\pi$: for example, if $\pi$ makes all matches as soon as possible, then $\m{N}^\pi$ is the set of states with at most one agent present.) We claim that $N^\pi$ is irreducible. This can be seen because for any time difference $\Delta t >0$ and any state $n \in \m{N}^\pi$, the empty state can be reached with positive probability within the next $\Delta t$ time periods, if all current agents abandon and no new ones arrive. Meanwhile, from the zero state, each state in $\m{N}^\pi$ is reachable by definition.

     We now use \citet[Theorem 2.8]{prieto2016uniform} to show that $N^\pi$ is ergodic. Let $V(n) = \sum_{i \in \mathcal{T}} n_i$, and $w(n) =1$. Fix any $c > 0$. Note that under any policy, the total number of agents in system $V(n)$ can only increase by one, and increases at a rate that is at most $\sum_{i \in \m{T}} \lambda_i$. It follows that for any state $n \in \m{N}^\pi$, the left side of \eqref{eq:stability} is at most $\sum_{i \in \m{T}} \lambda_i$, so if we take $b = c + \sum_{i \in \m{T}} \lambda_i$, then \eqref{eq:stability} is automatically satisfied for all $n \in C$.
     
     To ensure that \eqref{eq:stability} is satisfied for $n \not \in C$, we will define $C = \{n \in \mathcal{N}^\pi : V(n) < k\}$ for an appropriately chosen $k$. So long as $n^*$ is big enough, the expected change in $V$ from any state outside of $C$ will be less than $-c$. Specifically, it suffices to choose any $k > \frac{\sum_{i \in \m{T}} \lambda_i + c}{\min_{i \in \m{T}} \mu_i}$. The argument is as follows.
     \begin{align*}
        \sum_{n' \in \m{N}^\pi}q_{n, n'}(V(n') - V(n)) & =         \sum_{n' : V(n') > V(n)}q_{n, n'}(V(n') - V(n)) + \sum_{n' : V(n') < V(n)}q_{n, n'}(V(n') - V(n)) \\
        &\le \sum_{i \in \mathcal{T}} \lambda_i - \sum_{i \in \m{T}}n_i\mu_i\\
        &\le \sum_{i \in \mathcal{T}} \lambda_i - V(n)\min_i \mu_i\\
        & \le \sum_{i \in \mathcal{T}} \lambda_i - k\min_i \mu_i\\
        &< -c.
     \end{align*}
     The equality comes from splitting the sum into positive and negative terms. To obtain the first inequality, we upper bound the positive term as before: $V(n') - V(n) \le 1$, and the total rate of transitioning to a ``larger'' state is at most $\sum_{i \in \mathcal{T}} \lambda_i$. Meanwhile, we upper bound the negative term by noting that $V(n') - V(n) \le -1$, and the total rate of transitioning to a ``smaller'' state is at least the rate of abandonment $\sum_{i \in \m{T}}n_i\mu_i$. The second inequality follows by replacing each $\mu_i$ with $\min_i \mu_i$, the third follows because $n \not \in C$ and therefore $V(n) \ge k$, and the final inequality follows from our choice of $k$.
     
     We have shown that the conditions of \citet[Theorem 2.8]{prieto2016uniform} are satisfied, so $N^\pi$ is ergodic and thus has a unique stationary distribution. 
\end{proof}


\subsection{Proof of \Cref{prop:gammabound}}

\begin{proof}[Proof of \Cref{prop:gammabound}]
Fix a Markovian policy $\pi$, and let $N^\pi$ be sampled from its steady-state distribution, which is unique by \Cref{lemma:stationaryctmc}. Define 
\begin{equation}q_k^\pi =  \mathbb{P}(\sum_{i \in S} N_i^\pi = k)\end{equation}
to be the steady state probability that there are exactly $k$ agents with types belonging to $S$ in the system. Thus, we have 
\begin{equation} \frac{\E{\sum_{i \in S}N_i^\pi}}{\PP{\sum_{i \in S} N_i^\pi \ge 0}} =  \frac{\sum_{k \ge 1}k q_k^\pi}{\sum_{k \ge 1} q_k^\pi}  \label{eq:objective-q} \end{equation}

Define $\lambda_S = \sum_{i \in S} \lambda_i$. By assumption, $\mu_i = \mu$ for each $i \in S$. We claim that for $k \in \mathbb{N}$,
\begin{equation} q_k^\pi \lambda_S \ge q_{k+1}^\pi (k+1) \mu. \label{eq:balance-constraint}\end{equation}

To see this, define 
\begin{align}
\mathcal{N}_{k} &= \{\vec{n} \in \mathbb{N}^\mathcal{T} : \sum_{i \in S} n_i = k\} \\
\mathcal{N}_{\le k} &= \{\vec{n} \in \mathbb{N}^\mathcal{T} : \sum_{i \in S} n_i \le k\} \\
\mathcal{N}_{> k} &= \{\vec{n} \in \mathbb{N}^\mathcal{T} : \sum_{i \in S} n_i > k\}.
\end{align}
Consider the cut that separates the states $\mathcal{N}_{\le k}$ from the remaining states $\mathcal{N}_{> k}$. The balance equations imply that flow from $ \mathcal{N}_{\le k}$ to $\mathcal{N}_{> k}$ must equal flow from  $\mathcal{N}_{> k}$ to $\mathcal{N}_{\le k}$. Note that 

\begin{itemize}
\item From any state in $\mathcal{N}_{\le k-1}$, the rate of flow to $\mathcal{N}_{> k}$ is zero. \\ (We cannot have multiple simultaneous arrivals.)
\item From any state in $\mathcal{N}_{k}$, the rate of flow to $\mathcal{N}_{> k}$ is at most $\lambda_S$.\\ (This is the rate at which agents of some type $i \in S$ arrive to the system.)
\item From any state in $\mathcal{N}_{k+1}$, the rate of flow to $\mathcal{N}_{\le k}$ is at least $(k+1)\mu$. \\(This is the rate at which agents of some type $i \in S$ depart the system.)
\end{itemize}
It follows that 
\[ q_{k}^\pi \lambda_S \ge {\tt Flow}\,\, \mathcal{N}_{\le k} \to \mathcal{N}_{> k} = {\tt Flow}\,\, \mathcal{N}_{> k} \to \mathcal{N}_{\le k}  \ge q_{k+1}^\pi (k+1)\mu.  \]

Inspired by \eqref{eq:objective-q} and \eqref{eq:balance-constraint}, we consider the problem of choosing non-negative real numbers $\{q_k\}_{k \ge 1}$ to maximize 
\[\max_{q_k}  \frac{\sum_{k \ge 1}k q_k}{\sum_{k \ge 1} q_k}\]
subject to the constraints
\begin{equation} q_{k+1} \le q_k \frac{\lambda_S}{(k+1) \mu}. \label{eq:balance-constraint-2} \end{equation}
This relaxes the original problem, as we no longer require that the $q_k$ represent steady-state probabilities induced by a policy $\pi$. Since the objective and constraints are homogeneous of degree one, we add the constraint that $q_1 = (\lambda_S/\mu) e^{-\lambda_{S}/\mu}$ without loss of generality. Furthermore, we fix $m \in \mathbb{N}$ and impose the constraint that $q_k = 0$ for $k > m$. 

For each of these optimization problems (indexed by $m$), \Cref{lemma:optbetak-2} states that the objective value is maximized when each constraint in \eqref{eq:balance-constraint-2} is tight: that is, $q_k = e^{-\lambda_S/\mu} (\lambda_S/\mu)^k/k!$ is a truncated density of a Poisson random variable with mean $\lambda_S/\mu$.

Clearly, the objective value is increasing in $m$ (as $m$ increases, we relax constraints), and as $m \rightarrow \infty$, we have that the objective value converges to $(\lambda_S/\mu)/(1-e^{-\lambda_S/\mu}) = 1/\gamma_S$. Since $q_k^\pi$ is feasible, it follows from \eqref{eq:objective-q} that 
\[\frac{\E{\sum_{i \in S}N_i^\pi}}{\PP{\sum_{i \in S} N_i^\pi \ge 0}} \le \frac{1}{\gamma_S}.\]
\end{proof}

\begin{lemma}
    \label{lemma:optbetak-2}
    For positive real numbers $\{ \beta_k \}_{k = 0}^{m-1}$, consider the following optimization problem.
    \begin{align}
         \max_{q_k \ge 0} h(q) = &\frac{\sum_{k=1}^m k q_k}{\sum_{k=1}^m q_k}\\
        \text{s.t } \quad &q_{k+1} \le \beta_k q_k  \,\, \forall k = 1, 2, \ldots, m-1\label{lemma:con:beta}\\
        & q_1 = \beta_0.
    \end{align}
    In any optimal solution, we have
    \begin{equation}
        q_k = \prod_{j=0}^{k-1} \beta_j  \quad \forall k \in 1,2, \ldots, m \label{lemma:con:q:beta:res}
    \end{equation}
\end{lemma}

\begin{proof}
First note that the feasible space is compact, and the function $h$ is continuous on this domain. Therefore, it attains its supremum.

We claim that any feasible solution $q$ in which the constraints in \eqref{lemma:con:beta} are not all tight cannot be optimal. From this, it follows that in the optimal solution, all of the constraints in \eqref{lemma:con:beta} must be tight, implying that \eqref{lemma:con:q:beta:res} holds.

Our proof will use the following observation: if $a, b, c, d$ are non-negative real numbers with $c, d > 0$ and $a/c < b/d$, then $(a+\omega b)/(c+\omega d)$ is increasing in $\omega$. There are many ways to see this; one is to note that 
\[ \frac{d}{d\omega} \frac{a+\omega b}{c+\omega d} = \frac{(c+\omega d)b - (a+\omega b)d}{(c+\omega d)^2} = \frac{bc - ad}{(c+\omega d)^2} = \frac{cd(b/d - a/c)}{(c+\omega d)^2}> 0.\]

Take a feasible solution $q$, and let $k$ be a constraint for which \eqref{lemma:con:beta} is not tight. That is, $q_{k+1} < q_k \beta_k$. We will construct a feasible solution $p$ with a strictly higher objective value. We consider the cases $q_{k+1} = 0$ and $q_{k+1} > 0$ separately.

If $q_{k+1} = 0$, then \eqref{lemma:con:beta} implies that $q_j = 0$ for all $j > k$. Choose $p_{k+1} \in (0,\beta_k q_k)$, and let $p_j = q_j$ for $j \ne k+1$. Then $p$ is clearly feasible. This choice of $p$ ensures that
\begin{equation}h(p) =  \frac{\sum_{j=1}^k j q_j + (k+1) p_{k+1}}{\sum_{j=1}^k q_j+ p_{k+1}}. \label{eq:first-fraction} \end{equation}
Let $a = \sum_{j=1}^k j q_j$, $b = (k+1)$, $c = \sum_{j=1}^k q_j$, and $d = 1$. Note that $a/c$ is at most $k$, as it is 
a weighted average of terms from $1$ through $k$. Meanwhile $b/d = k+1$. Therefore, we can apply our observation to conclude that \eqref{eq:first-fraction} is increasing in $p_{k+1}$. Since $h(q)$ is the result of evaluating \eqref{eq:first-fraction} with $p_{k+1} = 0$, it follows that $h(q) < h(p)$. In other words, $q$ is not optimal.

If $q_{k+1} > 0$, fix $\omega \in (1, \beta_k q_{k}/q_{k+1})$. Set $p_j = q_j$ for $j \le k$, and $p_j = \omega q_j$ for $j > k$. It is clear that this choice satisfies \eqref{lemma:con:beta} (that is, this choice of $p$ is feasible). Furthermore,
\begin{equation}h(p) =  \frac{\sum_{j=1}^k j q_j + \omega \sum_{j = k+1}^m j q_j}{\sum_{j=1}^k q_j+ \omega \sum_{j = k+1}^m q_j}.\label{eq:second-fraction} \end{equation}
Define $a = \sum_{j=1}^k j q_j$, $b = \sum_{j=k+1}^m j q_j$, $c = \sum_{j=1}^k q_j$, $d = \sum_{j=k+1}^m q_j$. Note that $a/c$ is at most $k$, as it is 
a weighted average of terms from $1$ through $k$. Meanwhile $b/d$ is at least $k+1$, as it is a weighted average of terms from $k+1$ through $m$. Therefore, we can apply our observation to conclude that \eqref{eq:second-fraction} is increasing in $\omega$. Since $h(q)$ is the result of evaluating \eqref{eq:second-fraction} with $\omega = 1$, it follows that $h(q) < h(p)$. In other words, $q$ is not optimal.
\end{proof}

\subsection{Proof of \Cref{lemma:balance:multiple}}
\label{proof:lemma:balance:multiple}
\begin{proof}
        Let $A_i(t) = |\mathcal{A}_i(t)|$ be the number of agents of type $i$ that have arrive by time $t$. Recall that $N_i^\pi(t)$ represents the number of agents of type $i$ in the system at time $t$ and $X_{ij}^\pi(t)$ is the number of $(i,j)$ matches made before time $t$. 
        Let $D_i^{\pi}(t)$ denote the number of agents $i$ that have abandoned the system (departed unmatched) by time $t$. Then for any time $t$ and any $i \in \m{T}$, we have the following balance equation:
    
        \begin{equation}
            D_i^{\pi}(t) + N_i^{\pi}(t) + \sum_{j \in \mathcal{T}} X^\pi_{ij}(t) + X_{ji}^\pi(t) = A_i(t). 
        \end{equation}
        This says that each agent who arrives before time $t$ has either abandoned, been matched with another agent, or is still in the system. We also know that
        \begin{equation}
            \E{D_i^{\pi}(t)} = \E{\mu_i\int_0^t N_i^\pi(s)ds} \label{eq:d-ineq}
        \end{equation}
        because for any greedy policy $\pi$, $D_i^{\pi}(t) - \mu_i\int_0^t N_i^\pi(s)ds$ is a martingale (see for example, \citet{pang2007martingale}). Taking expectations and limits we get:
        \begin{align}
            \lim_{t \to \infty} \frac{1}{t} \left( \E{D_i^{\pi}(t)} + \E{N_i^{\pi}(t)} + \sum_{j \in \mathcal{T}} \E{X^\pi_{ij}(t)} + \E{X^\pi_{ji}(t)} \right) = \lim_{t \to \infty} \frac{1}{t}\E{A_i(t)}
        \end{align}
        Now combine the following facts:
        \begin{itemize}
             \item Equation \eqref{eq:d-ineq} and \Cref{lemma:stationaryctmc}, which imply that $\frac{1}{t} \E{D_i^{\pi}(t)} = \frac{\mu_i}{t} \int_0^t N_i^{\pi}(s)ds \rightarrow \mu_i \E{N_i^{\pi}}$ as $t\rightarrow \infty$.
            \item $N_i^{\pi}(t)$ is at most the number of agents who arrive before $t$ and would abandon after $t$. As explained in the proof of \Cref{prop:offline-bound}, this is a Poisson with mean $\frac{\lambda_i}{\mu_i}(1-e^{-\mu_i t}) \le \frac{\lambda_i}{\mu_i}$. Therefore, $\frac{1}{t}\E{N_i^{\pi}(t)} \le \frac{\lambda_i/\mu_i}{t}\rightarrow 0$ as $t\rightarrow \infty$.
            \item The definition of $x^\pi$ in \eqref{def:xpi}.
            \item $\frac{1}{t}\E{A_i(t)}= \lambda_i$.
        \end{itemize}
        Together, these imply that \eqref{eq:balancegreedy1} holds:
        \begin{equation}
            \E{N_i^\pi} \mu_i + \sum_{j \in \mathcal{T}} x^\pi_{ij} + x_{ji}^\pi = \lambda_i. \label{eq:end-result}
        \end{equation}
        
        To establish \eqref{eq:messy}, fix $j \in \m{T}$. For any $S \in Prefixes(\succ_j)$ define 
        \begin{align}
            X_{Sj}^\pi(t) & = \sum_{i \in S} X_{ij}^\pi(t) \\
            N_S^\pi(t) & = \sum_{i \in S} N_i^\pi(t).
        \end{align}
        That is, $X_{Sj}^\pi(t)$ is the number of matches made between an arriving agent of type $j$ with agents of some type $i \in S$ before time $t$, and $N_S^\pi(t)$ is the number of agents of some type $i \in S$ present in the system at time $t$. 
                
        Any greedy policy can only match an arriving $j$ to some type $i \in S$ if such an agent is present in the system. Therefore, for any $S$, $X_{Sj}(t) - \lambda_j \int_0^t {\bf 1}(N_S^\pi(s) > 0) ds$ is a sub-martingale, from which it follows that
        \begin{equation}
            \lim_{t \to \infty} \frac{1}{t} \E{X_{Sj}^\pi (t)} \le \lim_{t\to \infty} \frac{1}{t}\lambda_j \int_0^t {\bf 1}(N_S^\pi(s) > 0) ds \label{eq:wxy}
        \end{equation}
        and thus \eqref{eq:messy} holds:
        \begin{equation}
            x_{Sj}^\pi \le \lambda_j \PP{N_S^\pi >0}. \label{eq:xyz}
        \end{equation}
        If $S$ is a prefix of $\succ_j$, then any time $N_S^\pi(t) > 0$, an arriving $j$ will be matched to some agent in $S$, and therefore $X_{Sj}(t) - \lambda_j \int_0^t {\bf 1}(N_S^\pi(s) > 0) ds$ is a martingale. It follows that \eqref{eq:xyz} holds with equality. Conversely, if $S$ is not a prefix of $\succ_j$, we claim that \eqref{eq:wxy} and \eqref{eq:xyz} hold with strict inequality. This can be seen by noting that there is a positive probability of being in states where $N_S > 0$ and yet the arrival of a type $j$ would not cause $X_{Sj}$ to increment. In particular, suppose that $S$ is not a prefix of $\succ_j$, and let $i$ be a type that satisfies one of the following conditions: either $i \in S$ and $i$ is not acceptable according to $\succ_j$, or $i \not \in S$ and $i$ is preferred to some type in $S$. With positive probability, only agents of type $i$ are present. (This is because the empty state occurs with positive probability, and from this state, we transition at rate $\lambda_i$ to a state where only one type $i$ agent is present. Additionally, regardless of the policy, we cannot leave this state until the agent of type $i$ abandons the system or another agent arrives.) In any state where only type $i$ is present, $N_S^\pi > 0$ and yet $X_{Sj}$ would not increment upon the arrival of a type $j$. Hence, \eqref{eq:xyz} holds with strict inequality.         

                   
        To prove the final claim, note that if \eqref{eq:messy} holds with equality, we know that $S$ is a prefix of $\succ_j$ and in particular that each $i \in S$ is an acceptable match for $j$. It follows that $x_{ij} > 0$ for all $i \in S$, because for each $i \in S$, states where only type $i$ agents are present occur with positive probability (as explained in the preceding paragraph). In these states, $\pi$ will make an $(i,j)$ match upon arrival of an agent of type $j$, implying that the rate of $(i,j)$ match formation $x_{ij}^\pi$ is strictly positive.
    \end{proof}

\subsection{Proof of \Cref{lemma:suitablestructure}}
\label{proof:lemma:suitablestructure}

We start by presenting results related to a function that is closely related to the constants $\gamma_S$.

\begin{lemma}
    \label{lemma:hcvx}
    \label{lemma:usingcvxh}
    Define the function $h(x) = \frac{x}{1-e^{-x}}$. Then 
    \begin{itemize}
        \item $h$ is increasing and convex in the domain $(0, \infty)$. 
        \item For any $x_1, x_2, x_3 > 0$ we have that,
    \begin{equation}
        h(x_1 + x_2) +  h(x_1 + x_3) < h(x_1) + h(x_1 + x_2 + x_3).
    \end{equation}
    \end{itemize}
\end{lemma}
\begin{proof}
    {\bf First fact:}
    Taking the first and second derivate,
    \begin{align}
        \frac{d h}{d x} &= \frac{e^{x} (e^{x} - x-1)}{(e^x - 1)^2}\\
        \frac{d^2 h}{d x^2} &= \frac{e^{m} (e^{x} (x-2) + x + 2)}{(e^x - 1)^3}
    \end{align}
    we can verify that both are strictly positive in the domain.

    {\bf Second fact:}
    We are going to use the convexity of $h$. Choose $\alpha = \frac{x_3}{x_2+x_3}$ and note that $\alpha\cdot x_1 + (1-\alpha)(x_1 + x_2 + x_3) = x_1+x_2$. Similarly, if we choose $\beta = \frac{x_2}{x_2+x_3}$ we get that $\beta \cdot x_1 + (1-\beta)(x_1 + x_2 + x_3) = x_1+x_3$.
    By convexity of $h$ we have:
    \begin{align*}
        \alpha h(x_1 )+(1-\alpha)h(x_1 + x_2 + x_3) &> h(x_1 + x_2)\\
        \beta h(x_1 )+(1-\beta)h(x_1 + x_2 + x_3) &> h(x_1 + x_3)
    \end{align*}
    Adding these inequalities and noting that $\alpha + \beta = 1$ gets the stated result.
\end{proof}

An important intermediate result shows the nested structure of sets in $\m{S}^j$ of feasible solutions of $\plpm$. 

\begin{lemma}
    \label{lemma:nestedsets}
    Let $\m{M}$ be a set of matches and $({\bf n}, {\bf x}, \bm{\psi})$ be a feasible solution of $\plpm$. Let $j \in \m{T}$ be a type and $S,S' \subseteq T(\m{M}, j)$ be distinct sets of types such that
    \begin{align}
        \sum_{i \in S} x_{ij} &= \lambda_j\gamma_{S} \sum_{i \in S} n_{i} \label{eq:s-tight}\\
        \sum_{i \in S'} x_{ij} &= \lambda_j\gamma_{S'} \sum_{i \in S'} n_{i} \label{eq:s'-tight}
    \end{align}
    and $x_{ij} >0$ for $i \in S \cup S'$. Then either $S \subseteq S'$ or $S' \subseteq S$. 
\end{lemma}
\begin{proof}
    For any set $T$, we let $x_{Tj} = \sum_{i \in T} x_{ij}$ and $n_{Tj} = \sum_{i \in T} n_i$. We show the result by contradiction: if neither $S$ nor $S'$ contains the other, then constraint  \eqref{eq:match-rate-lp-glb} is violated for $S \cup S'$:
        \begin{equation}
        x_{S \cup S'j} > \lambda_j \gamma_{S \cup S'} n_{S \cup S'}.
    \end{equation}
    
    Suppose that neither $S \subset S'$ nor $S' \subset S$ and therefore $S \cap S' \not = S$ and $S \cap S' \not = S'$. Then multiplying \eqref{eq:s-tight} by $\frac{\gamma_{\{S\cup S'\}}}{\gamma_{S}}$ and \eqref{eq:s'-tight} by $\frac{\gamma_{\{S\cup S'\}}}{\gamma_{S'}}$ and adding them yields:
    \begin{align}
        \frac{\gamma_{\{S\cup S'\}}}{\gamma_{S}} x_{Sj} + \frac{\gamma_{\{S\cup S'\}}}{\gamma_{S'}}x_{S'j} &= \lambda_j\gamma_{\{S\cup S'\}}(n_{S \setminus S'} + n_{S' \setminus S} + 2 n_{S \cap S'}) \label{eq:partway}
    \end{align}
        Because $S \cap S' \subseteq T(\m{M}, j)$, we know from constraint \eqref{eq:match-rate-lp-glb} that $x_{S \cap S'} \le \lambda_j \gamma_{S \cap S'} n_{S \cap S'}$ and therefore, $-x_{S \cap S'} \frac{\gamma_{S\cup S'} }{\gamma_{S \cap S'}} \ge -\lambda_j \gamma_{S \cup S'} n_{S \cap S'}$. Adding these terms to \eqref{eq:partway} and expanding $x_{Sj} = x_{S \cap S' j} + x_{S \setminus S' j}$ and $x_{S'j} = x_{S \cap S' j} + x_{S' \setminus S j}$ yields
    \begin{align} \label{eq:bigmess}
        \left(\frac{\gamma_{S\cup S'} }{\gamma_S} + \frac{\gamma_{S\cup S'} }{\gamma_{S'}} - \frac{\gamma_{S\cup S'} }{\gamma_{S \cap S'}} \right) x_{S\cap S' j} + \frac{\gamma_{S \cup S'}}{\gamma_S} x_{S\setminus S' j} + & \frac{\gamma_{S \cup S'}}{\gamma_{S'}} x_{S'\setminus S j} \\ & \ge \lambda_j \gamma_{S \cup S'} (n_{S \setminus S'} + n_{S' \setminus S} + n_{S \cap S'}). \nonumber \\
        & = \lambda_j \gamma_{S \cup S'} n_{S \cup S'}. \nonumber
    \end{align}


    Apply \Cref{lemma:usingcvxh} with $x_1 = \sum_{i \in S \cap S'} \lambda_i/\mu_i $, $x_2 = \sum_{i \in S\setminus S'} \lambda_i/\mu_i $, $x_3 = \sum_{i \in S'\setminus S} \lambda_i/\mu_i $ to get
    \begin{equation}
        \gamma_{\{S\cup S'\}} \left(\frac{1}{\gamma_{S}} + \frac{1}{\gamma_{S'}} - \frac{1}{\gamma_{S\cap S' }} \right) < 1. \label{eq:l1}
    \end{equation}
    Furthermore, because of \Cref{lemma:hcvx} (increasing $h$) we have: 
    \begin{align}
        \frac{\gamma_{\{S\cup S'\}}}{\gamma_{S}}&<1 \label{eq:l2} \\
        \frac{\gamma_{\{S\cup S'\}}}{\gamma_{S'}}&<1. \label{eq:l3}
    \end{align}
    We conclude that 
    \begin{align*}
        x_{S \cup S'j} & = x_{S \cap S'j} +  x_{S\setminus S'j} + x_{ S' \setminus Sj} \\
        &  >   \left(\frac{\gamma_{S\cup S'} }{\gamma_S} + \frac{\gamma_{S\cup S'} }{\gamma_{S'}} - \frac{\gamma_{S\cup S'} }{\gamma_{S \cap S'}} \right) x_{S\cap S' j} + \frac{\gamma_{S \cup S'}}{\gamma_S} x_{S\setminus S' j} + \frac{\gamma_{S \cup S'}}{\gamma_{S'}} x_{S'\setminus S j} \\
        & \ge \lambda_j \gamma_{S \cup S'} n_{S \cup S'},
    \end{align*}
    where the strict inequality follows from \eqref{eq:l1}, \eqref{eq:l2}, \eqref{eq:l3} and the fact that $x_{ij} > 0$ for all $i \in S\cup S'$, and the second inequality follows from \eqref{eq:bigmess}. Because $S \cup S' \subseteq T(\m{M}, j)$, this contradicts feasibility of $({\bf n},{\bf x})$.
\end{proof}


Finally, we identify the possible domain of the variables $\{n_i\}_{i \in \m{T}}$:

\begin{lemma}
    \label{lemma:ppoisglb}
    Let $\m{M}$ be a set of matches and $({\bf n}, {\bf x}, \bm{\psi})$ be a feasible solution of $\plpm$, then $n_i \in (0, \lambda_i/\mu_i]$ for all $i \in \m{T}$.
\end{lemma}
\begin{proof}
    First note that from \eqref{eq:balance-lp-glb} we get that for any $i \in \m{T}$, $n_i \mu_i \le \lambda_i$ and therefore
    \begin{equation}
        n_i \le \lambda_i/\mu_i \label{eq:96} 
    \end{equation}
    Now note that if $n_i = 0$, then LHS of \eqref{eq:balance-lp-glb} is merely $\sum_{j \in T(\m{M}, i)} x_{ji}$, which by \eqref{eq:match-rate-lp-glb} and \eqref{eq:96} has to be strictly less than $\lambda_i$ yielding a contradiction.
\end{proof}

We now use these results to show \Cref{lemma:suitablestructure}.
\begin{proof}
    The number of constraints in $\plpm$ is $|\m{T}| + \sum_{j \in \m{T}} (2^{|T(\m{M},j)|}-1)$ and the number of variables is $|\m{T}| + |\m{M}| +  \sum_{j \in \m{T}} (2^{|T(\m{M},j)|}-1)$. 
    A basis $\m{B}$ is a subset of $|\m{T}| + \sum_{j \in \m{T}}( 2^{|T(\m{M},j)|}-1)$ variables.  Because $({\bf n}, {\bf x}, {\bf \psi})$ is a basic feasible solution, there must exist a basis $\m{B}$ such that all variables outside of $\m{B}$ are zero. \Cref{lemma:ppoisglb} states that in any feasible solution of $\plpm$, all $n_i$ are strictly positive, and therefore these variables must be in the basis $\m{B}$. We therefore conclude that for any basic feasible solution, the number of $x_{ij}$ variables in the basis is equal to the number of $\psi_{Sj}$ variables excluded from the basis.  
    
    Fix $j \in \m{T}$. By \Cref{lemma:nestedsets}, we know that among the sets $\m{S}^j = \{S : \psi_{Sj} = 0\}$, there is one that contains all others. Call this set $\m{L}^j$. By the definition of suitability, for each $i \in \m{L}^j$ we must have $x_{ij} > 0$ (and thus $x_{ij} \in \m{B}$). Therefore, the number of $i$ such that $x_{ij}$ is included in the basis $\m{B}$ is at least $|\m{L}^j|$. Furthermore, the number of sets $S$ such that $\psi_{Sj}$ is {\em not} in $\m{B}$ is at most the number of sets $S$ such that $\psi_{Sj} =0$, which is at most $|\m{L}^j|$, since \Cref{lemma:nestedsets} says that these sets must be nested. From our conclusion in the previous paragraph, it follows that the number of sets $S$ such that $\psi_{Sj} = 0$ must be exactly $|\m{L}^j|$. Because these sets are nested, there must be exactly one of each size $k$ for each $k\in \{1,2, ...,|\m{L}^j|\}$.
    \end{proof}
    
    \subsection{Proof of \Cref{prop:offline-bound}}

\label{proof:prop:offline-bound}
\begin{proof}
    For any subsets $S, S'\subseteq \mathcal{T} $ we want to see how many agents of type $i$ arrive and either observe an agent of some type $j \in S$ or depart after the arrival of some agent of type $j' \in S'$. This will give an upper bound on the number of matches of the form presented in \eqref{eq:offlinematchrate}. We are going to use a result similar to the one presented in \citep{wolff1982poisson} that says that on any finite interval, the expected number of arrivals who observe a given state is equal to the arrival rate times the probability of being in said state. In our case the events of interest are (i) finding an agent of type $j\in S$ that has not left unmatched by the time of the arrival and (ii) the arrival of an agent of type $j' \in S'$ before the unmatched departure of the arriving agent.

    We start by analyzing the probability that for any time $t$ there is an agent of type $j \in S$ in the system. Define the following random variable that counts the number of agents of type $j$ in the system at time $t$,
    \begin{equation}
        Z_j(t) = |\{b \in \mathcal{A}_j(t) : d_b > t\}|.
    \end{equation}
    Because $A_j(t) = |\m{A}_j(t)| \sim Poisson(\lambda_j t)$ if we condition on the number of arrivals, the arrival time $t_b \sim Unif(0,t)$ for all $b \in \mathcal{A}_j(t)$. Therefore, conditioned on the number of arrivals of type $j$, we have that for any agent $b \in \mathcal{A}_j(t)$, the probability that they haven't departed by time $t$ is:
    \begin{align}
        \mathbb{P}(d_b >t | A_j(t) = k) &= \int_0^{t} \mathbb{P}(d_b >t | t = s, A_j(t) = k) \mathbb{P}(t_b =s | A_j(t) = k)ds\\
        &=  \int_0^{t} \mathbb{P}(d_b-t_b >t-t_b | t_b = s, A_j(t) = k) \mathbb{P}(t_b =s | A_j(t) = k)ds\\
        &=  \int_0^{t} \mathbb{P}(d_b-t_b >t_b | t_b = s, A_j(t) = k) \mathbb{P}(t_b =s | A_j(t) = k)ds\\
        &=  \int_0^{t} e^{-s\mu_j} \frac{1}{t}  ds \\
        & = \frac{1-e^{-\mu_jt}}{t\mu_j}
    \end{align}
    we used $d_b - t_b \sim Exp(\mu_j)$ for the fourth equality and the fact that for $t_b \sim Unif(0,t)$ we have that $\mathbb{P}(t_b = s) = \mathbb{P}(t_b = t-s)$ for the third (or equivalently $t_b \sim Unif(t, 0)$). 
    Because each agent's departure time is independent we have that:
    \begin{align}
        Z_j(t) \sim Poisson\left( \frac{\lambda_j}{\mu_j}\left( 1-e^{-\mu_jt}\right) \right)
    \end{align} 
    Let $Z_{S}(t)$ be the random variable that counts the number of agents of type $j \in S$ at time $t$,
    \begin{equation}
        Z_S(t) = \sum_{j \in S} Z_j(t)
    \end{equation}
    then
    \begin{equation}
        Z_S(t) \sim Poisson\left(\sum_{j \in S} \frac{\lambda_j}{\mu_j}\left( 1-e^{-\mu_jt}\right) \right)
    \end{equation}

    Now we are going to analyze the probability that for time $t$ an agent of some time $j \in S'$ arrives before some exponentially distributed time $\Delta \sim Exp(\mu_i)$. To this end, for any $\Delta>0$ define the following random variable that counts the number of agents of some type $j \in S'$ that arrive after $t$ and before $t+ \Delta$:
    \begin{equation}
        Y_{S'}(t, \Delta) = |\{b \in \mathcal{A}(t+\Delta) : t < t_b, i_b \in S'  \}|
    \end{equation}
    If $\Delta \sim Exp(\mu_i)$ then because the arrival rate of agents follow a Poisson process we get that 
    \begin{equation}
        \PP{Y_{S'}(t,\Delta) = 0} = \frac{\mu_i}{\mu_i + \sum_{j \in S'} \lambda_j}
    \end{equation}
    Now that we have studied the arrival of future and past arrivals we want to know the fraction of agents of type $i$ that arrive when either a type $j \in S$ is present or depart after the arrival of some agent $j \in S'$.  These agents are given by the following random variable,
    \begin{equation}
        W(t) = \{a \in \mathcal{A}_i(t) : Z_S(t_a) + Y_{S'}(t_a, d_a-t_a)  >0,  \}.
    \end{equation}

    
    Note that if  $\Delta$ is an independent random variable then for any $t' > t >0$ the following random variables are independent: $Z_S(t), A_i(t') - A_i(t), Y_{S'}(t',\Delta)$. 
    
    We can approximate $W(t)$ with the following sequence of random variables. Let $\Delta_k \sim Exp(\mu_i)$ be a family of independent random variables, then:
    \begin{align}
        W_n(t) = \sum_{k=0}^{n-1} \mathbb{I} \left (Z_S \left(\frac{kt}{n} \right) + Y_{S'}\left(\frac{(k+1)t}{n}, \Delta_k\right) > 0 \right) \left[A_i\left(\frac{(k+1)t}{n} \right) - A_i\left(\frac{kt}{n} \right) \right]
    \end{align}
    Because $W_n(t) \le A_i(t)$ and $A_i(t)$ is integrable we can use the Dominated convergence theorem 
    \begin{align}
        E[W(t)] &= E\left[ \lim_{n \to \infty} W_n(t) \right] \\
        &=   \lim_{n \to \infty} E\left[ W_n(t) \right] \\
        &=  \lim_{n \to \infty}  \lambda_i \frac{t}{n} \sum_{k=0}^\infty  \mathbb{P} \left (Z_S \left(\frac{kt}{n} \right) + Y_{S'}\left(\frac{(k+1)t}{n}, \Delta_k\right) > 0 \right)\\
        &= \lambda_i \int_0^t \mathbb{P} \left (Z_S(s) + Y_{S'}(s,\Delta_1) > 0 \right)ds\\
        &= \lambda_i \int_0^t 1-\mathbb{P} \left (Z_S(s) + Y_{S'}(s,\Delta_1) = 0 \right)ds\\
        &= \lambda_i \int_0^t1-\mathbb{P} \left (Z_S(s) = 0 \right)\mathbb{P} \left (Y_{S'}(s,\Delta_1) = 0 \right)ds
    \end{align}
    Where in the last equality we used the independence between $Z_{S}(t)$ and $Y_{S'}(t, \Delta_1)$. 
    We can use the previously calculated distributions of $Z_S(t)$ and $Y_{S'}(t,\Delta)$ for $\Delta \sim Exp(\mu_i)$ to get:
    \begin{align}
        E[W(t)] &= \lambda_i \int_0^t1-\mathbb{P} \left (Z_S(s) = 0 \right)\mathbb{P} \left (Y_{S'}(s,\Delta) = 0 \right) ds\\
        &= \lambda_i \int_0^t1-\mathbb{P} \left (Y_{S'}(s,\Delta) = 0 \right) \prod_{j \in S} \mathbb{P} \left (Z_j(s) = 0 \right)\\
        &\le \lambda_i \int_0^t 1 - \frac{\mu_i}{\mu_i + \sum_{j \in S'} \lambda_j} \prod_{j \in S}  e^{- \lambda_j /\mu_j}\\
        & = \lambda_i t \left(1-\frac{\mu_i}{\mu_i + \sum_{j \in S'} \lambda_j} e^{-\sum_{j \in S} \lambda_j /\mu_j} \right)
    \end{align}
    This is true for any sets $S, S' \subseteq \mathcal{T}$ type $i \in \mathcal{T}$ and so we can bound the rate of matches using $E[W(t)]/t$ which matches the constraints described in \eqref{eq:offlinematchrate2}. It is easy to see that $\frac{1}{t} \E{X_{ij}^\pi}(t) \ge 0$ since this is a counting process and therefore \eqref{eq:posrate2} follows.
\end{proof}

\subsection{Proof of \Cref{thm:no-better-half}}
\label{appendix:no-better-half}


    We start by rewriting the linear program that \citet{aouad2022dynamic} (Claim 3) use to upper bound any online policy, using our notation. 
    \begin{align}
        \textup{LP}^{\textup{ON}}(\m{I}) =& \quad  \max_{{\bf n}, {\bf x}} \sum_{i,j \in \m{T}} r_{ij} x_{ij} \label{online:lp}\\
        \text{ s.t }& \quad   n_i \mu_i +\sum_{j \in \m{T}}x_{ij} + x_{ji} = \lambda_i \quad \forall i \in \m{T} \label{eq:s1}\\
        & \quad x_{ij} \le  n_i \lambda_j\quad \forall i,j \in \m{T} \label{eq:xp-ub} \\
        & \quad x_{ij}, n_i \ge 0 \quad \forall i,j \in \m{T}
    \end{align}
    
    \begin{lemma}[\citet{aouad2022dynamic}, Claim 3.]
        \label{lemma:onlineup}
        For any instance $\m{I}$ and any deterministic stationary policy $\pi$ we have $\textup{LP}^{\textup{ON}}(\m{I})\ge V(\pi, \m{I})$.
    \end{lemma}

    \begin{example} \label{example:hard} 
    There two types $\m{T} = \{1, 2\}$, with $\lambda_1 = 1, \mu_1 = \mu_2 = \mu$, $r_{11} = 2 + \mu$, $r_{12} = r_{21} = 1$, and $r_{22} = 0$.
    \end{example}
    
    \begin{lemma}
        \label{lemma:onlinefor2}
        For any value of $\lambda_2$ in the instance from \Cref{example:hard}, we have $\textup{LP}^{\textup{ON}}(\m{I}) \le 1$.
    \end{lemma}
    \begin{proof}
        Take a feasible solution $({\bf n}, {\bf x})$ of $\textup{LP}^{\textup{ON}}(\m{I})$. The objective value of this solution is 
    \[(\mu+2)x_{11} + x_{12} + x_{21} \le \mu n_1 + 2 x_{11} + x_{12} + x_{21},\]
    where we have used that \eqref{eq:xp-ub} implies $x_{11} \le n_1 \lambda_1 = n_1$.  But the right side above is exactly the left side of \eqref{eq:s1} for $i = 1$, which must be equal to $\lambda_1 = 1$. In other words, for any feasible solution, the objective value is $1 - \mu(n_1 - x_{11})$.

    Now that we have an upper bound for any non-anticipatory policy we proceed to show that having future information could lead to a better objective. non-Markovian policies could do better. 
    For this, consider the following policy (which uses future arrival and departure times). Let $t_k$ be the time of the $k^{th}$ type $1$ arrival, and $d_k$ the time at which this arrival will depart unmatched. 
    
    If $d_k > t_{k+1}$ and $d_{k+1} < t_{k+2}$, the policy matches $k$ to $k+1$. Otherwise, it attempts to match $i$ to a type $2$.
    
    Note that the constraints on $d$ and $t$ imply that this does define a feasible matching: $k$ is matched to $k+1$ only if they coincided, and it is never the case that $k$ is matched to both $k+1$ and $k-1$. Denote this policy by ADJ (short for {\em adjacent}, because we only match adjacent pairs of agents of type $1$).
    \begin{lemma}
        \label{lemma:adj}
        In the instance from \Cref{example:hard}, as $\lambda_2 \to \infty $ we have $V(\textup{ADJ}, \m{I}) \rightarrow \frac{2\mu^2 + 2\mu + 1}{(1+\mu)^2}$.    
    \end{lemma}
    
    \begin{proof}
        Let ${\bf x}^{\text{ADJ}}$ be the long-term rate of matches under policy ADJ. Then under policy ADJ, an arriving agent of type $1$ has a probability of $\frac{1}{1+\mu}\left(1 - \frac{1}{1+\mu} \right) = \frac{\mu}{(1+\mu)^2}$ of being the first arrival in a $(1,1)$ match. Therefore, the rate of $(1,1)$ matches under policy ADJ is $x_{11}^{\text{ADJ}} = \frac{\mu}{(1+\mu)^2}$. Becuase $\lambda_2 \to \infty$ no agent of type $1$ goes to waste, and if they don't match with a type $1$ they will match with a type $2$ and therefore,
        \[ x_{12}^{\text{ADJ}} + x_{21}^{\text{ADJ}} = 1 - 2\frac{\mu}{(1+\mu)^2} = \frac{\mu^2 + 1}{(1+\mu)^2}\]
        Thus, the rate of reward of this policy is,
        \[V(\text{ADJ}, \m{I}) = (2+\mu) \cdot x_{11}^{\text{ADJ}}  + 1 \cdot (x_{12}^{\text{ADJ}} + x_{21}^{\text{ADJ}}) = \frac{\mu(2+\mu) + \mu^2 + 1}{(1+\mu)^2} = 2 - \frac{2\mu+ 1}{(1+\mu)^2} \]
    \end{proof}

   Jointly, \Cref{lemma:onlinefor2} and \Cref{lemma:adj} imply that for any online policy $\pi$ in the instance $\m{I}$ from \Cref{example:hard}, as $\lambda_2 \rightarrow \infty$ we have
    \begin{align*}
        \frac{V(\pi, \m{I})}{\offopt} \le \frac{\textup{LP}^{\textup{ON}}(\m{I})}{V(\text{ADJ}, \m{I})} = \frac{1}{2 - \frac{2\mu+ 1}{(1+\mu)^2}}.
    \end{align*}    
    Because the previous expression goes to $1/2$ as $\mu \to \infty$, \Cref{thm:no-better-half} follows: for any $\epsilon > 0$ there must exist some $\mu$ such that $\frac{1}{2 - \frac{2\mu+ 1}{(1+\mu)^2}} < 1/2 + \epsilon$, as stated.
    \end{proof}

\section{Additional Results}
\subsection{Competitive Ratio of $1/2$ when $|\m{T}| = 2$}
\label{sec:app:twotypes}

\begin{theorem}
    \label{thm:two-types}
    Let $\m{I}$ be an instance where $\m{T}=\{1, 2\}$ and $r_{12} = r_{21}$. Then 
    \[V(\alg,\m{I}) \ge \frac{1}{2} \offopt.\]
\end{theorem}

\begin{proof}
We have established that for {\em any} instance $\m{I}$, if $\m{M}$ is the set returned by \Cref{algo:suitability}, then
\[ \plpm \ge \plpt \ge \frac{1}{2} \offrbrel \ge \frac{1}{2} \offopt.  \]
Therefore, all that we need to show is that 
\[ V(\alg, \mathcal{I}) \ge \plpm.\]

Let $({\bf n}, {\bf x}, \bm{\psi})$ be the optimal suitable solution associated with $\m{M}$, so that $\pi = \alg$. By \Cref{prop:plplowerbound}, it suffices to show that \eqref{eq:alpha-gamma} holds for all sets $S$ such that $\psi_{Sj}$ for some $j$.

For any set $S$ of size one, \Cref{prop:gammabound} implies that \eqref{eq:alpha-gamma} holds. The only set of size greater than one is $S = \{1, 2\}$. Suppose that $\psi_{Sj} = 0$ for some $j \in \{1, 2\}$. By \Cref{lemma:value-based} this optimal solution is consistent with the values  ${\bf v}$ from the optimal solution of $\dlpm$. If $1 \succ_2 \emptyset$ then $r_{12} - v_1 - v_2 \ge 0$ and because $r_{12} = r_{21}$ we also have that $r_{21} - v_1 - v_2 \ge 0$. If $r_{21} - v_1 - v_2 > 0$ then $2\succ_1 \emptyset$. The case when  $r_{21} - v_1 - v_2 = 0$ is not interesting as this implies that $z_{Sj} = 0$ for all $S \subseteq \{1,2 \}$  and $j \in \{1,2 \}$ and therefore $v_1 = v_2 = 0$ and this can only happen if $r_{ij} = 0$ for all $i,j \in \{1,2\}$. From this we can conclude that if $\pi$ makes $(1,2)$ matches then it also makes $(2,1)$ matches.  Therefore, for corresponding greedy policy of $({\bf n}, {\bf x}, \bm{\psi})$ it follows that 
\begin{align*}\mathbb{P}(N_1^\pi + N_2^\pi  > 0)  & = \mathbb{P}(N_1^\pi > 0) + \mathbb{P}(N_2^\pi  > 0) \\
& \ge \gamma_1 n_1^\pi + \gamma_2n_2^\pi \\
& \ge \gamma_{12}( n_1^\pi + n_2^\pi).
\end{align*}
The first inequality applies \eqref{eq:alpha-gamma} for $S = \{1\}$ and $S = \{2\}$, while the second uses the fact that $\gamma_{12} \le \min(\gamma_1, \gamma_2)$. These inequalities establish that \eqref{eq:alpha-gamma} holds for $S = \{1, 2\}$. 
\end{proof}

\subsection{Competitive Ratio of $(e-1)/(2e)$ for Bipartite Instances}
\label{sec:app:aouadguarantee}

In this section we present a proof that the result from \citet{aouad2022dynamic} can be combined with our \Cref{prop:offline-bound} and \Cref{prop:factor2} to get a new guarantee for bipartite instances.  We rewrite the result from \citet{aouad2022dynamic} that we will use.

\begin{lemma}[\citet{aouad2022dynamic}]
    \label{lemma:aouad}
    Let $({\bf n}, {\bf x})$ be a feasible solution of $\textup{LP}^{\textup{ON}}(\m{I})$ as defined in \eqref{online:lp} then there exists a Markovian policy $\pi$ such that, 
    \begin{equation}
        V(\pi, \m{I}) \ge \kappa \left(1 - 1/e \right) \sum_{i,j \in \m{T}} r_{ij} x_{ij}
    \end{equation}
    where $\kappa = 1/2$ if $\m{I}$ is a bipartite instance and $\kappa = 1/4$ otherwise. 
\end{lemma}

From this, we obtain the following result.


\begin{theorem}
    For any instance $\m{I}$ there exists a Markovian policy $\pi$ such that,
    \begin{equation}
        V(\pi, \m{I}) \ge \frac{\kappa}{2} \left(\frac{e-1}{e} \right)\offopt
    \end{equation}
    where $\kappa = 1/2$ if $\m{I}$ is a bipartite instance and $\kappa = 1/4$ otherwise. 
\end{theorem}
\begin{proof}
Let $({\bf n}, {\bf x},{\bf \psi})$ be a feasible solution of $\plpt$. Then it immediately follows that $({\bf n}, {\bf x})$ is a feasible solution of $\textup{LP}^{\textup{ON}}(\m{I})$. This is because \eqref{eq:xp-ub} can be thought of as a relaxation of \eqref{eq:match-rate-lp-glb} where only singleton sets are considered, and where each $\gamma_i$ is increased to 1. Then \Cref{lemma:aouad} states that there exists a Markovian policy $\pi$ such that
    \begin{align*}
        V(\pi, \m{I}) &\ge \kappa \left(\frac{e-1}{e} \right) \sum_{i,j \in \m{T}} r_{ij} x_{ij}  \\
        &= \kappa \left(\frac{e-1}{e} \right) \plpt \\
        &\ge  \frac{\kappa}{2} \left(\frac{e-1}{e} \right) \offrbrel \\
        &\ge \frac{\kappa}{2} \left(\frac{e-1}{e} \right) \offopt,
    \end{align*}
    where the final inequalities follow from \Cref{prop:factor2} and \Cref{prop:offline-bound}, respectively. The result for general instances yields a factor of $(e-1)/(4e)$ which is below the $1/8$ proved by \citet{collina2020dynamic}, but to the best of our knowledge the result for bipartite instances is new.
\end{proof}


\subsection{Assumptions on $\mu$ are Necessary for \Cref{prop:gammabound}.} 
\label{sec:app:alphasmaller}

Fix a small $\epsilon > 0 $ and consider the following instance: $\lambda_1 = \mu_1 = \epsilon$,  $\lambda_2 = \mu_2 = 1$ and $\lambda_3 = 1/\epsilon$ ($\mu_3$ can be any value). Consider a policy $\pi$ which operates as follows. Never match type 1. If there is a type 1 in the system, do not match type 2. Otherwise (if $N_1 = 0$), match type 2 with type 3. We claim that $\PP{N_1^\pi + N_2^\pi > 0} <  \gamma_{\{1,2\}} (\E{N^\pi_1} + \E{N^\pi_2})$.

Note that because we make no matches involving type 1, we have
\begin{align*}
    \PP{N^\pi_1 > 0} & =1 - e^{-\lambda_1/\mu_1} =  1 - e^{-1}\\
    \E{N^\pi_1} &= \lambda_1/\mu_1 = 1.
\end{align*}
Epochs where $N_1 = 0$ have an average duration of $1/\epsilon$, while those where $N_1 > 0$ have an average duration of $(e-1)/\epsilon$. During epochs where $N_1 = 0$, all type $2$ agents in the system are matched almost immediately, so $\PP{ N_2^\pi  > 0 | N^\pi_1 = 0}$ is $\mathcal{O}(\epsilon)$. During epochs where $N_1 > 0$, the number of type 2 agents behaves as an $M/M/\infty$ queue with a mixing time that does not depend on $\epsilon$ (is $\mathcal{O}(1)$). Therefore, $\E{N^\pi_2 | N^\pi_1 > 0} \approx \lambda_2/\mu_2 = 1$. It follows that 
\begin{align*}
    \PP{N^\pi_1 + N_2^\pi > 0} & =  \PP{N^\pi_1 > 0} + \PP{N_1^\pi = 0}\PP{ N_2^\pi  > 0 | N^\pi_1 = 0} = 1 - e^{-1} + \mathcal{O}(\epsilon) \\
    \E{N^\pi_2} &= \PP{N_1^\pi >0}\E{N^\pi_2 | N^\pi_1 >0} + \PP{N_1^\pi = 0}\E{N^\pi_2 | N^\pi_1 = 0}  \\
    &= 1 - e^{-1} + \mathcal{O}(\epsilon).
\end{align*}
Therefore,
\begin{align*}
    \PP{N^\pi_1 + N^\pi_2 > 0} & \approx 1 - e^{-1} = 0.632 \\
    \gamma_{\{1,2\}}(\E{N^\pi_1}+\E{N^\pi_2}) & \approx \frac{1 - e^{-2}}{2} (2-e^{-1}) \approx 0.706.
\end{align*}



\end{document}